\documentclass{acta}

\usepackage{supertabular,lscape,epsfig}
\usepackage{amssymb}
\usepackage{amsmath}
\usepackage{url}
\usepackage{color}
\usepackage{xspace}
\usepackage{bm}

\newcommand{\mathbmss}[1]{\bm{\mathsf{#1}}}
\usepackage{natbib}

\SetPages{0}{0}

\SetVol{70}{2020}

\begin{document}

\begin{Titlepage}
\Title{High-velocity moving groups in the Solar neighborhood in Gaia DR2}
\Author{B~a~l~u~e~v, R.~V., \quad S~h~a~i~d~u~l~i~n, V.~Sh., \quad and V~e~s~e~l~o~v~a, A.~V.}
{Saint Petersburg State University, 7--9 Universitetskaya Emb., St Petersburg 199034, Russia\\
e-mail: r.baluev@spbu.ru}

\Received{March 02, 2020}
\end{Titlepage}

\Abstract{We use an improved wavelet analysis technique to reconstruct the $(U,V,W)$
velocity distribution for $\sim 250000$ stars from Gaia DR2, residing in the solar
neighborhood of $200$~pc. The 2D wavelet transforms for three bivariate distributions
$(U,V)$, $(U,W)$, and $(V,W)$ were investigated. Though most of currently known
(low-velocity) stellar moving groups are densely overlapped in these diagrams, our analysis
allowed to detect and disentangle about twenty statistically significant 3D groups of stars
with high velocities. Most of them appear new. We also discuss the issue of correct noise
thresholding in the wavelet transform and highlight the importance of using a global rather
than local statistic for that. Using of a local significance measure may lead to an
overstated statistical confidence for individual patterns due to the effect of multiple
testing.}{Moving groups, Milky Way kinematics, Gaia DR2, Solar neighborhood, statistical
wavelet analysis.}

\section{Introduction}
Stellar moving groups in the solar neighbourhood have been studied for a long time. In the
20th century a detailed study was performed by Olin Eggen starting in the 1960s
\citep[e.g.][]{Eggen65,Eggen96}, in particular, he investigated the structure of the Sirius
and Hyades moving groups \citep{Eggen1960a,Eggen1960b}. The Hyades group itself along with
the Ursa Major group has been known since the 19th century \citep[see][]{Proctor1869}
according to data on alignment of the velocities of several stars. Over the 20th and 21st
centuries  a variety of stellar streams and moving groups including Tucana--Horologium,
$\beta$ Pictoris, TW Hydrae, AB Doradus and other have been discovered and investigated
(see reviews in \citealt{Torres08-book,Gagne18,LeeSong19}).

A variety of methods was created for detection of moving groups and classification of
individual stars into groups. A series of algorithms BANYAN~I \citep{Malo13}, BANYAN~II
\citep{Gagne14}, BANYAN~$\Sigma$ \citep{Gagne18} as well as LACEwING \citep{Riedel17} take
into account not only the space velocities UVW, but also the Galactic coordinates XYZ.
Complex algorithms also include constraints based on distances and colors/magnitudes.

In a number of studies a detailed analysis of structures on the $U$--$V$ plane was carried
out. \citet{Michtchenko18} proposed that the stellar moving groups in the solar
neighborhood could be explained  by the spiral arms perturbations. Authors proposed that
Coma-Berenices and Hyades-Pleiades groups are located inside of the corotation radius and
the Sirius group is related to several overlapping outer Lindblad resonances.
\citet{Monari19} analyzed the ridge-like structures in local velocity space and discussed
structures that could be formed by the resonances of the large bar of the Galaxy. Authors
showed that several ridges can be related to resonances with the bar and two of them can be
associated with the Hercules moving group in local part of velocity space.

\citet{Ramos18} performed a decomposition of the distribution of approximately 5~million
stars in the $V_R$--$V_{\phi}$ plane with the wavelet transformation in a number of regions
near the Sun. Authors found several long structures on the $V_R$--$V_{\phi}$ plane with
approximately constant azimuthal velocity. Several structures follow lines of nearly
constant energy and resemble the Sirius group. Hyades or Hercules probably follow lines of
nearly constant angular momentum. Such different dynamical characteristics may indicate
different origins of the structures: Sirius stream may be related to phase-mixing
processes, but Hyades/Hercules may be formed as a resonant structure due to perturbation of
bar and/or the spiral arms. Authors also observed several candidates moving groups and some
of them are not related to already known groups.

The additional criteria such as the photometrical data or chemical composition are used to
determine the membership probabilities for stars. \citet{CochraneSmith19} compiled samples
of late-type dwarf stars for twelve moving groups and analyzed the loci of ensembles of
stars on two-color diagram (a (FUV--$B$) color index plotted against ($B$--$V$)). Authors
found the relationship between the photometric index based on (FUV--$B$) color and the ages
of late-type dwarf stars. They suppose that addition of FUV photometry would be useful for
membership determination of late-F, G, and K dwarf stars.

\citet{LeeSong18,LeeSong19} created models for nearby stellar young moving groups and
presented lists of confirmed members and model parameters of groups. Authors emphasized the
difference of membership probabilities from BANYAN series of methods; for example,
according to \citep{LeeSong19} HR 8799 belongs to the $\beta$ Pictoris moving group, but in
other studies this star is considered as a member of Columba group.

It is important to note the discussion of the status of stellar groups within $100$~pc by
\citet{Mamajek15}. Author stated that the Ursa Major, Hyades, Coma Ber and $\eta$~Cha groups
are real star clusters and their stellar densities exceed the density of the local part of
Galactic disc, the association TW Hya was called a well-characterized group of
approximately 3~dozen stars, and the Tuc-Hor group could be an ensemble of several
subgroups. The Columba, Carina, 32~Ori groups and $\chi^1$~For cluster were reported to be
the likely physical groups. Mamajek stated that the Argus, Oct-Near, Her-Lyr, Castor, IC
2391 Supercluster are unlikely physical groups; Her-Lyr, Castor and IC~2391 may probably be
considered as streams.

As we can see, many stellar moving groups are already known in the literature. They are
summarized below in Appendix~\ref{sec_kmg}. However, this is an ongoing discussion still
containing multiple controversies and ambiguities. In this work we undertake an attempt to
perform a systematic search for \emph{statistical} moving groups in the UVW-space, based on
the Gaia DR2 \citep{Gaia18,Gaia18kin}. We intend to use the statistical wavelet analysis
algorithm presented initially in \citep{Baluev18a} for univariate (1D) distributions and
recently extended to the bivariate (2D) analysis \citep{BalRodShai19}. This method has
certain improvements in comparison with wavelet analysis previously used by other authors
\citep{Skuljan99,Romeo08,BB10,BB16,BB18,Ramos18}. In particular, it has an important accent
on the correct determination of statistical significance. In fact, this work also
represents a new field-test for this 2D algorithm, in addition to the asteroid families
search made by \citet{BalRod20}.

The structure of the paper is as follows. In Section~\ref{sec_samples} we describe how our
statistical samples are constructed based on the Gaia data. In Section~\ref{sec_wan} we
describe several points of our wavelet analysis algorithm and discuss some conventions
related to the formal criteria behind a statistical cluster or a group. In
Section~\ref{sec_mg} we present our main results of the analysis.

\section{Data samples}
\label{sec_samples}
We exported a portion of Gaia DR2 from \textit{Vizier} database, containing stars with
known radial velocities. This resulted in more than $7$ million records. Then stars with
parallax lower than $5$ mas (those outside the heliocentric radius of $200$ pc) were
removed. Inside this solar neighborhood stars with big relative parallax uncertainties
(greater then $20\%$) were also removed. This left $352000$ records. Then the
transformation to galactic velocities $U$, $V$, $W$ was performed using procedure described
in \textit{Gaia Data Release 2 Documentation}
(\url{https://gea.esac.esa.int/archive/documentation/GDR2/}).

Briefly, for each star we can compute local basis:
\begin{equation}
\mathbf p=
\left[\begin{array}{c}
-\sin\alpha\\
\cos\alpha\\
0
\end{array}\right]
,\qquad
\mathbf q=
\left[\begin{array}{c}
-\cos\alpha\sin\delta\\
-\sin\alpha\sin\delta\\
\cos\delta
\end{array}\right]
,\qquad
\mathbf r=
\left[\begin{array}{c}
\cos\alpha\cos\delta\\
\sin\alpha\cos\delta\\
\sin\delta
\end{array}\right],
\end{equation}
where $\alpha$, $\delta$ are the right ascension and declination, respectively. Then
velocity is
\begin{equation}
\mathbf v=\frac\kappa\varpi\mu^*_\alpha\mathbf p+\frac\kappa\varpi\mu_\delta\mathbf q+v_r\mathbf r.
\label{vel}
\end{equation}
Here $\kappa=4.740\,470\,446$ expresses the astronomical unit in km\,yr\,s$^{-1}$, $\varpi$
is parallax, $\mu^*_\alpha$ is proper motion in right ascension with $\cos\delta$ factor,
$\mu_\delta$ is proper motion in declination, $v_r$~is radial velocity. We can obtain
galactic velocities by rotation of velocity $\mathbf v$ to galactic frame with orthogonal
matrix $A$:
\begin{equation}
\left[\begin{array}{c}
U\\
V\\
W
\end{array}\right]
=\mathrm A\mathbf v,
\end{equation}
$$
\mathrm A=\left[\begin{array}{ccc}
-0.0548755604162154&-0.8734370902348850&-0.4838350155487132\\
+0.4941094278755837&-0.4448296299600112&+0.7469822444972189\\
-0.8676661490190047&-0.1980763734312015&+0.4559837761750669
\end{array}\right].
$$

Finally, we removed the stars with too big velocity uncertainties greater than
$3$~km\,s$^{-1}$. For that purpose we calculated a velocity covariance matrix for each
star, using the uncertainties provided by Gaia DR2. We apply the so-called delta method.
First, the velocity Jacobian is
\begin{equation}
\mathbmss J=\frac{\partial\mathbf v}{\partial\mathbf u}=\left[\frac{\partial\mathbf v}{\partial\alpha},\frac{\partial\mathbf v}{\partial\delta},\frac{\partial\mathbf v}{\partial\varpi},\frac{\partial\mathbf v}{\partial\mu^*_\alpha},\frac{\partial\mathbf v}{\partial\mu_\delta},\frac{\partial\mathbf v}{\partial v_r}\right], \quad
\mathbf u = \left[\alpha,\delta,\varpi,\mu^*_\alpha,\mu_\delta,v_r\right]^{\mathrm T}
\end{equation}
and it is a $3\times 6$ matrix. Here,
\begin{align}
\frac{\partial\mathbf v}{\partial\alpha}&=\frac\kappa\varpi\mu^*_\alpha\frac{\partial\mathbf p}{\partial\alpha}+\frac\kappa\varpi\mu_\delta\frac{\partial\mathbf q}{\partial\alpha}+v_r\frac{\partial\mathbf r}{\partial\alpha}\,,\qquad
\frac{\partial\mathbf v}{\partial\delta}=\frac\kappa\varpi\mu_\delta\frac{\partial\mathbf q}{\partial\delta}+v_r\frac{\partial\mathbf r}{\partial\delta}\,,
\nonumber\\
\frac{\partial\mathbf v}{\partial\varpi}&=-\frac\kappa{\varpi^2}\mu^*_\alpha\mathbf p-\frac\kappa{\varpi^2}\mu_\delta\mathbf q,\qquad
\frac{\partial\mathbf v}{\partial\mu^*_\alpha}=\frac\kappa\varpi\mathbf p,\qquad
\frac{\partial\mathbf v}{\partial\mu_\delta}=\frac\kappa\varpi\mathbf q,\qquad
\frac{\partial\mathbf v}{\partial v_r}=\mathbf r,
\nonumber\\
\frac{\partial\mathbf p}{\partial\alpha}&=
\left[\begin{array}{c}
-\cos\alpha\\
-\sin\alpha\\
0
\end{array}\right]
,\qquad
\frac{\partial\mathbf q}{\partial\alpha}=
\left[\begin{array}{c}
\sin\alpha\sin\delta\\
-\cos\alpha\sin\delta\\
0
\end{array}\right]
,\qquad
\frac{\partial\mathbf r}{\partial\alpha}=
\left[\begin{array}{c}
-\sin\alpha\cos\delta\\
\cos\alpha\cos\delta\\
0
\end{array}\right],
\nonumber\\
\frac{\partial\mathbf q}{\partial\delta}&=
\left[\begin{array}{c}
-\cos\alpha\cos\delta\\
-\sin\alpha\cos\delta\\
-\sin\delta
\end{array}\right]
,\qquad
\frac{\partial\mathbf r}{\partial\delta}=
\left[\begin{array}{c}
-\cos\alpha\sin\delta\\
-\sin\alpha\sin\delta\\
\cos\delta
\end{array}\right].
\end{align}

The small difference of the velocity $\mathbf v$ is expressed through a linear
approximation as
\begin{equation}
\Delta\mathbf v = \mathbmss J \Delta \mathbf u.
\end{equation}
Then the $3\times 3$ variance-covariance matrix is approximated as
\begin{equation}
\mathrm{Var}(\mathbf v) = {\mathbb E} \Delta\mathbf v {\Delta\mathbf v}^{\mathrm T} \simeq \mathbmss J \mathrm{Var}(\mathbf u) \mathbmss J^{\mathrm T}.
\end{equation}
The matrix $\mathrm{Var}(\mathbf v)$ defines an ellipsoid in $\mathbb R^3$, and we adopt
its semi-major axis $\sigma_v$ (or the largest matrix eigenvalue) as an indicative
characteristic of the velocity uncertainty.

\begin{figure}[tb]
\includegraphics[width=\textwidth]{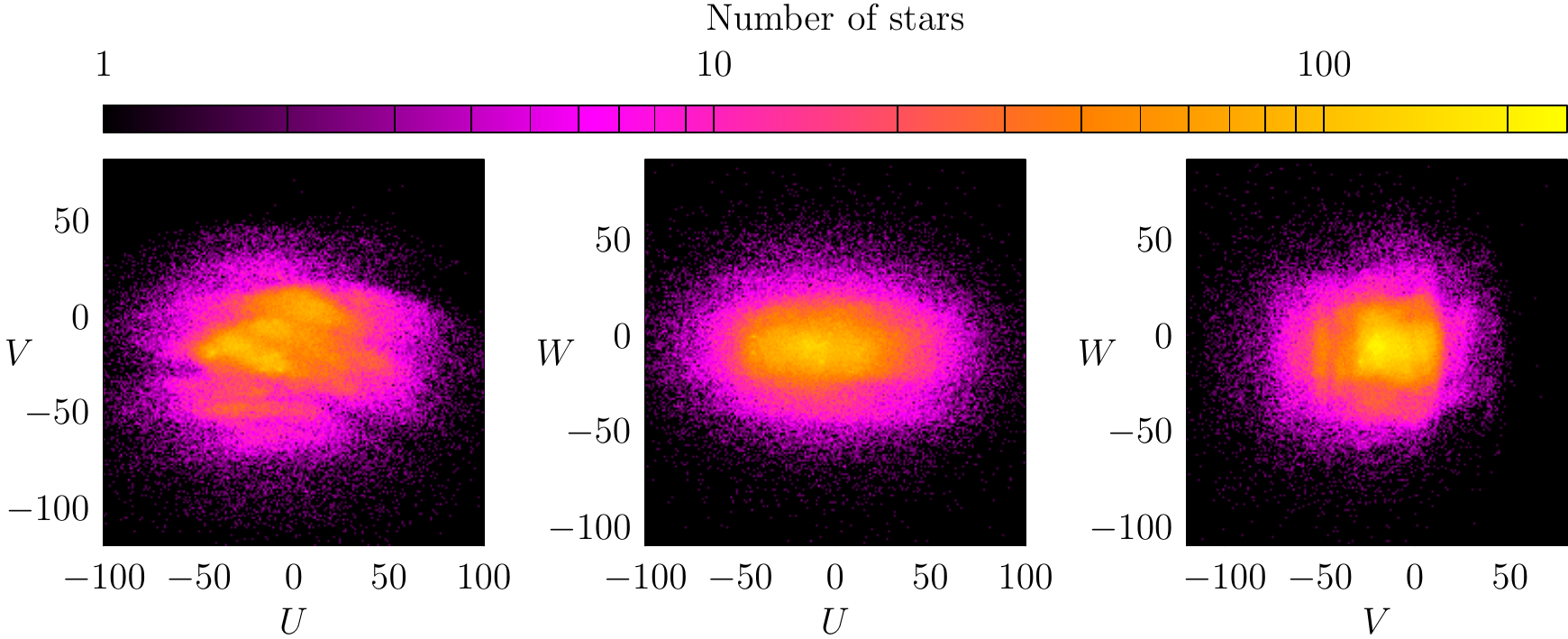}
\FigCap{\label{fig:dataset} Three bivariate velocity distributions: $UV$--plane,
$UW$--plane, $VW$--plane (from left to right, respectively).}
\end{figure}

Our primary sample contains only stars with $\sigma_v<3$~km\,s$^{-1}$.
Fig.~\ref{fig:dataset} shows three 2D histograms of this sample. Its size is $N=245499$
sources. We also considered a smaller subsample with the uncertainty limit below
$1$~km\,s$^{-1}$ which resulted in only $N=45879$ sources.

We also considered additional indicators available in Gaia DR2 that could be helpful in
removing unreliable sources from our samples. In particular, we investigated the so-calles
excess astrometric noise of a source, $\epsilon_i$. This is an additional error term added
quadratically to the observational uncertainies. Mathematically, it is analogous to the
`jitter' known well in radial velocity surveys \citep{Wright05,Baluev08b}. It appears if
the empiric scatter of observation is significantly larger then their expected
uncertainties. Multiple reasons can cause such a disagreement, e.g. (i) additional noise
sources, or (ii) various model inaccuracies. In the first case, this $\epsilon_i$ is
already taken into account in our $\sigma_v$ through the uncertainies in $\mathbf u$. The
effect of $\epsilon_i$ decreases in such a case as $1/\sqrt{N_{\rm obs}}$, where $N_{\rm
obs}$ in number of observations of a source. The second case indicates some systematic
errors of the astrometric model, and one its possible cause is the presence of an unseen
companion (astrometric binary). In such a case its effect would not decrease as
$1/\sqrt{N_{\rm obs}}$, because it is not a random quantity. Given that, the uncertainties
of $\mathbf u$ may appear understated as they implicitly imply such a factor through
astrometric model fitting. Hence, the values of $\varpi$ and $\mu$ might then appear
distorted more than expected, and so would the derived UVW components. We therefore decided
to roughly estimate the effect of $\epsilon_i$ for each source, under the assumption that
it is a systematic error rather than random noise.

This can be done as follows. Given the formula~(\ref{vel}), the astrometric part of the
spatial velocity vector is $\kappa \mu/\varpi$. In this treatment the `jitter' $\epsilon_i$
represents some characteristic systematic bias of each observation. Such a bias may distort
either the parallax $\varpi$, resulting in a velocity error $\sim \kappa \epsilon_i
\mu/\varpi^2$, or the proper motion $\mu$, generating a error of $\sim
\kappa\epsilon_i/(\varpi T)$, where $T$ is the observation time span. Combining these
quantities into a sum-of-squares metric, we obtain
\begin{equation}
\delta v_{\rm sys} \sim \frac{\kappa \epsilon_i}{\varpi} \sqrt{\frac{1}{T^2}+\left(\frac{\mu}{\varpi}\right)^2}.
\end{equation}
This quantity can be used as a measure of possible systematic errors that may appear e.g.
because of unclassified astrometric companions of a source. We computed $\delta v_{\rm
sys}$ for each object in our samples, and filtered out those that had it above $3$~km/s (or
$1$~km/s, respectively). Setting $T=5$~yr, we found just about $80$ such sources in the
larger sample, and only $1$ in the smaller one. These potentially problematic sources were
spread over the UVW space without clear concentration, so they are unlikely to affect our
statistical analysis, even if their spatial velocity is inaccurate.

\section{Wavelet analysis and treatment of statistical clusters}
\label{sec_wan}
We are going to use the wavelet analysis method described in \citep{Baluev18a} and
\citep{BalRodShai19}. It is based on the continuous wavelet transform (CWT):
\begin{equation}
Y(a,b) = \int\limits_{\mathbb R^n} f(x) \psi\left(\frac{x-b}{a}\right)\, dx,
\label{cwt}
\end{equation}
where $f(x)$ is the probability density function (PDF) under study, $n=1$ or $n=2$ is the
dimension of the task (of the random variable $x$ and of the shift parameter $b$). The
scale parameter $a$ is always a scalar quantity, and $\psi(t)$ is an analysing wavelet. The
wavelet is assumed to be radially symmetric for $n=2$, so it is actually a function of just
the length of its argument vector.

The analysis algorithm involves the following stages:
\begin{enumerate}
\item Estimation of the CWT~(\ref{cwt}) based on the input random sample.
\item Computing the associated noise thresholding (or goodness-of-fit) statistic relative
to some comparison (null) model.
\item Construction of the most economic (in terms of the number of wavelet coefficients)
PDF model based on the iterative CWT inversion, until all deviations decrease below the
noise thresholds.
\end{enumerate}
The algorithm involves optimized minimum-noise wavelets and inversion kernels. Also, we
emphasize especially that it relies on the extreme value distribution when performing the
noise thresholding. This is different from many other works that use, basically, a single
value distribution for that goal. We advocate that the latter approach is inadequate for
this task, because it ignores the fact that multiple values of $Y(a,b)$ (multiple wavelet
coefficients) are tested simultaneously in a single analysis. In such a case it is
necessary to include a statistical correction for the effect of multiple testing: the
probability to make a statistical mistake among many values is much larger than for just a
single value. Otherwise, the noise thresholds would appear inadequately tight and a lot of
noisy wavelet coefficients may be wrongly declared as significant. This issue is broadly
analogous to the effect of unknown frequency in the periodogram analysis
\citep{Baluev08a,Suveges14}.

Unfortunately, unjustified neglection of this issue is shared between multiple wavelet
analysis works regarding the stellar statistics. In particular, we believe that
\citep{Skuljan99} and more recently \citep{Ramos18} overestimated the number of detectable
structures. In the latter work they appeared so numerous that authors decided to reduce
their number by joining them into branches, and discarded those that appeared outside of
any branch. However, the true underlying issue could be just an overstated significance
that appeared due to multiple testing of numerous wavelet coefficients.

However, in this work we would like to make additional focus on the interpretation issue
instead of just significance. This is a further development of the ideas presented in the
1D analysis by \citet{Baluev18b}, in particular in what concerns the formal definition of a
\emph{statistical cluster}. The last term is not strict: one may adopt quite different
criteria of what PDF structures should be attributed to isolated clusters of objects.
Although there is no universally applicable formal definition of a cluster, we still need
to set up some formalized convention or at least a guide how do we distinguish a cluster
from other PDF anomalies, especially in the 2D case.

The wavelet analysis inherently implies such a formalized definition. Let us first note
that our algorithm may use a wavelet function of one of the following forms:
\begin{equation}
\psi(t) = \left\{ \begin{array}{ll} \varphi'(t)\; \mathrm{or}\; \varphi''(t), & n=1, \\ \Delta \varphi(t), & n=2, \end{array} \right.
\label{psi}
\end{equation}
where $\Delta\varphi$ is the Laplace operator and $\varphi$ is a `generating function',
which represents a bell-like smoothing kernel. This generating function is determined in an
optimal way to reduce the noise, and is different for each wavelet, see
\citep{Baluev18a,BalRodShai19}.

The definition~(\ref{psi}) implies that $Y(a,b)$ is either a smoothed first or second
derivative of $f(x)$, if $n=1$, or a smoothed Laplacian of $f(x)$, if $n=2$. In this work,
we do not use the case with the first derivative (the WAVE/WAVE2 wavelets), since it
usually appeared not so informative regarding practical results. Therefore, the wavelet
analysis here implies that our primary objective is either the second derivative $f''(x)$
or the Laplacian $\Delta f(x)$:
\begin{equation}
Y(a,b) = a^2 \int\limits_{\mathbb R^n} \Delta f(x) \varphi\left(\frac{x-b}{a}\right) dx
 \stackrel{a\to 0}{\sim} a^{n+2} \Delta f(b) \int\limits_{\mathbb R^n} \varphi(t) dt.
\label{laplacian}
\end{equation}
This formula is obtained through integrating~(\ref{cwt}) by parts.

Hence, the CWT of this type is simply an analogue of the statistical kernel density
estimator (KDE), but (i) applied to $\Delta f(x)$ rather than to $f(x)$, and (ii) involving
all possible smoothing scales $a$ at once. It is important that $\Delta f(x)$, or just
$f''$ in the 1D case, plays a crucial role here. In fact, the CWT simply estimates them via
smoothing.

The particular shape of the smoothing kernel $\varphi$ was seeked from the optimality
condition minimizing noise in the CWT estimate. The optimal kernel is different for $n=1$
and $n=2$, resulting in the so-called CBHAT and 2DOPT wavelets, respectively
\citep{Baluev18a,BalRodShai19}. Notice that these wavelets have a qualitatively similar
shape to the classical MHAT wavelet, but still are quite different. The MHAT wavelet (the
second derivative of a Gaussian) is not suitable in tasks of this type because it generates
too much non-Gaussian skewed noise in the CWT.

\begin{figure}[tb]
\includegraphics[width=0.8\textwidth]{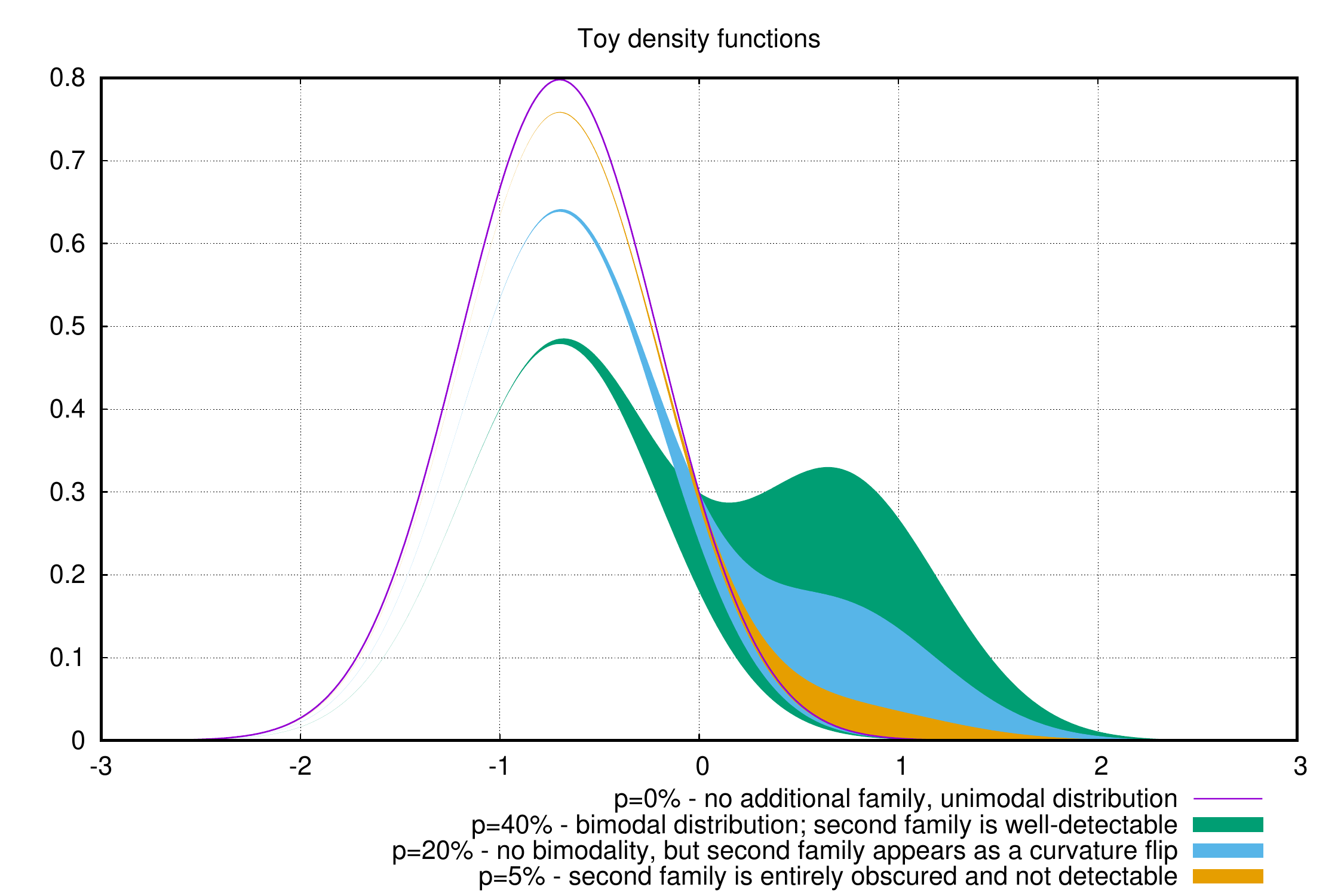}\\
\includegraphics[width=0.8\textwidth]{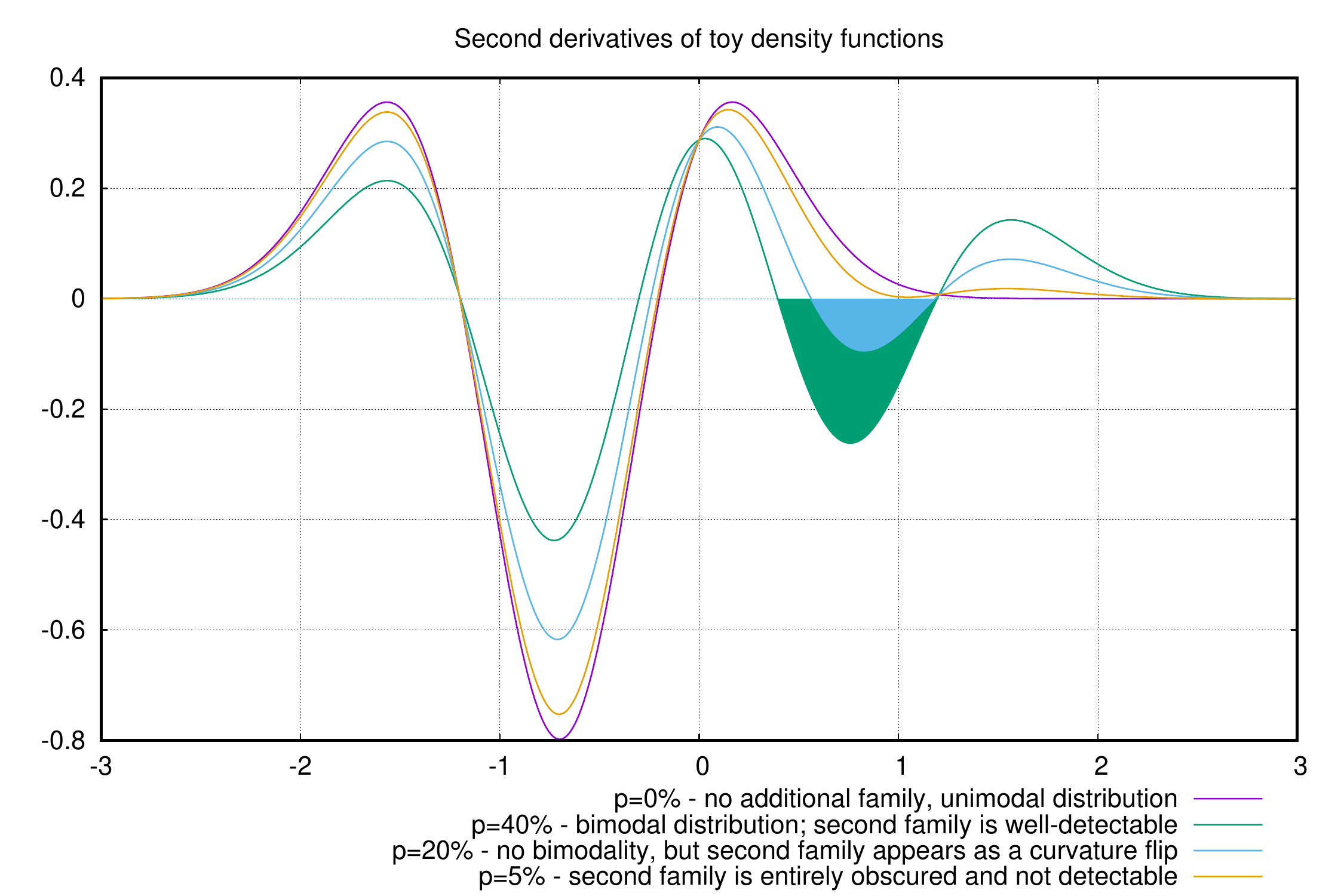}
\FigCap{\label{fig:group} Illistration of the formalized definition of a statistical
cluster (or a group).}
\end{figure}

Now let us consider how to define a cluster. The 1D case is illustrated in
Fig.~\ref{fig:group}. We consider a few toy PDF models there, each is a superposition of
two Gaussians that model two displaced statistical families of objects. The Gaussians
always have the same width (variance) and constant positions, but different relative
contribution. The relative contribution of the secondary family varies from $p=0$ to
$p=0.4$.

One might argue that the PDF should simply be bimodal to ensure that there are two families
instead of just one, but it appears that such a requirement is excessively strong. As we
can see, the PDF is bimodal only in the most prominent case $p=0.4$, but the second family
still can be easily noticed even if there is no bimodality. In particular, we can see clear
hint of an inhomogeneous structure for $p=0.2$. This appears because the curvature of the
graph changes the sign. It suddenly becomes upward-convex at the secondary family (as it
was near the primary peak).

This observation suggests that the second derivative $f''$ provides a good formal criterion
of a statistical cluster. Formally, a cluster can be defined as zone where $f''(x)$ is
negative. If there is no additional sign changes in $f''$, as for $p=0.05$, then the second
family becomes very uncertain. It appears just like a PDF excess that can be interpreted in
various alternative ways. This definition of a cluster via the sign changes of $f''$ is
basically a formalized version of the guides presented in \citep{Baluev18b}.

The 2D clusters can be determined in a similar way. In this case we consider the Laplacian
$\Delta f(x)$ which is estimated by our wavelet analysis via the kernel smoothing.
Formally, a 2D cluster can be defined as any disconnected domain where $\Delta f<0$ (the
PDF is convex). Importantly, this core domain would necessarily be surrounded by a
counter-signed `concavity ridge'. If there is no such a ridge then the domain is not
disconnected (not isolated), and the PDF pattern cannot be reliably identified as a
cluster.

Unfortunately, even these formal criteria appeared incomplete for our practical goals. They
are constructed only for the cases with $n=1$ or $n=2$, but in actuality we deal with at
least $n=3$ parameters $(U,V,W)$. We currently do not have a working 3D wavelet analysis
algorithm, because it requires additional theory work related to the noise significance
thresholds, and additional computing optimizations. We have to deal with 2D marginalized
(projected) PDFs of the 3D velocity distribution. This may reveal useful results too, but
our objective criterion of a cluster is then diluted. In particular, the overlapping effect
may take place, when two clusters partly merge in some 2D projection, but they can be
distinguished in the third dimension. In such conditions we adopt our cluster criteria as a
useful guiding, but we treat them more adaptively. For example, we additionally investigate
possible substructures within all isolated convexity domains.

\section{Disentangling moving groups detected by 2D wavelet analysis}
\label{sec_mg}
In this work we do not utilize the CWT~(\ref{cwt}) directly. This appears quite difficult
in the 2D case, because the CWT should then be considered as a function in 3D space.
Leaving this CWT at intermediate computing stages without detailed interpretation, we
finally consider the denoised PDF models reconstructed by our iterative matching pursuit
algorithm from the estimated CWT. These models represent the most economic models still in
a statistical agreement with the input sample \cite[see details in][]{BalRodShai19}.

If some structure appears in such a PDF model then it should necessarily pass the adopted
significance threshold in the CWT. In practice, however, the PDF may simultaneously contain
multiple structures with quite different magnitude (e.g. statistical clusters of different
richness). In particular, small-scale structures usually have smaller magnitude too. And
due to the poor contrast, the weaker patterns remain hardly seen in the PDF, even if they
are pretty significant in comparison with noise. In \citep{BalRod20} this issue was solved
by considering the Laplacian $\Delta f$ instead of the PDF model $f$ itself. The Laplace
operator highlights all smaller-scale structures, so such structures are seen well in
$\Delta f(x)$ regardless of the backgound (which is suppressed). Also, the Laplacian is
useful to outline statistical clusters, as described in Sect.~\ref{sec_wan}

\subsection{1D analysis}
\begin{figure}[tb]
\includegraphics[width=\textwidth]{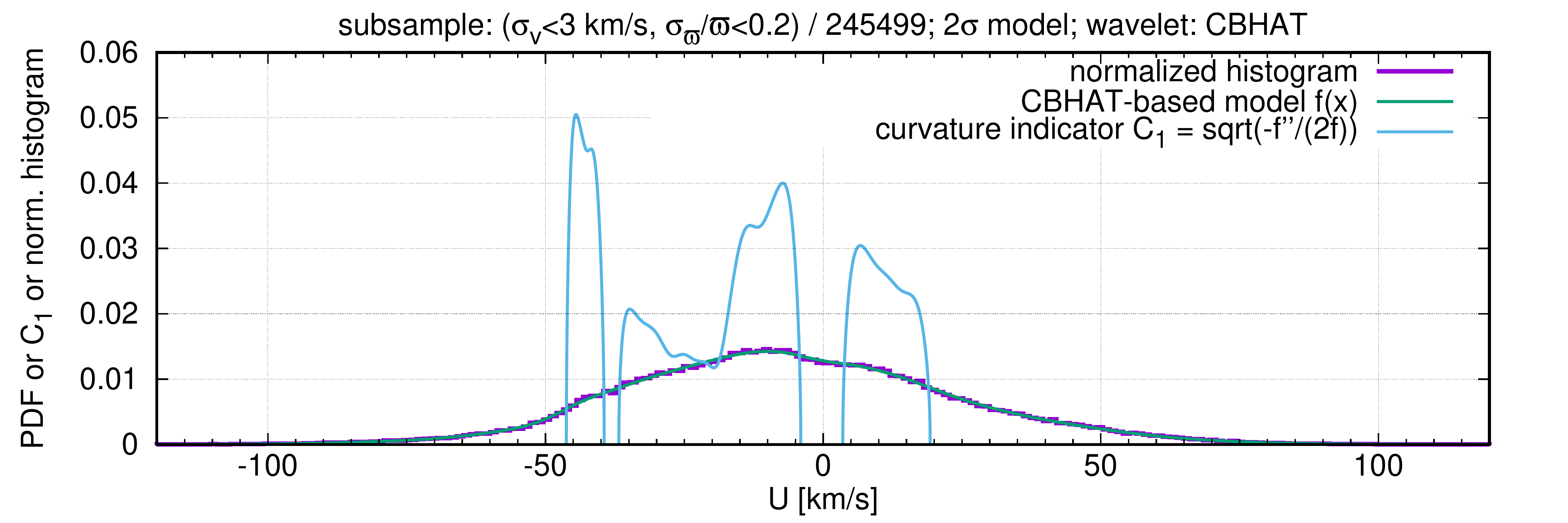}\\
\includegraphics[width=\textwidth]{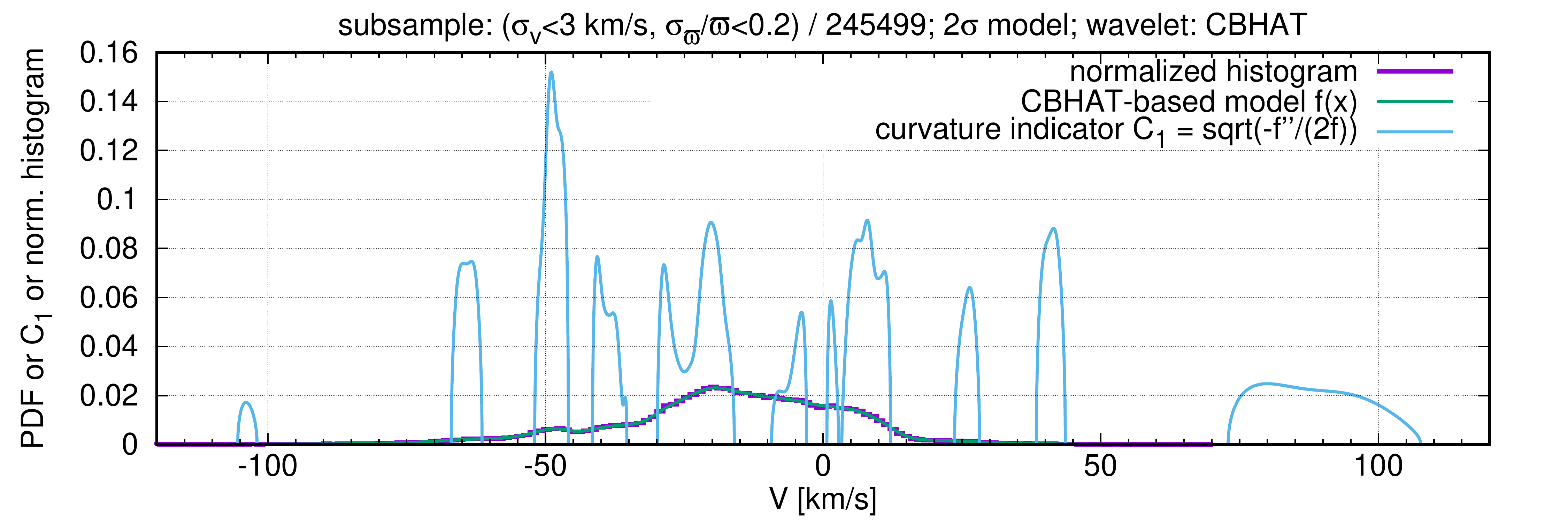}\\
\includegraphics[width=\textwidth]{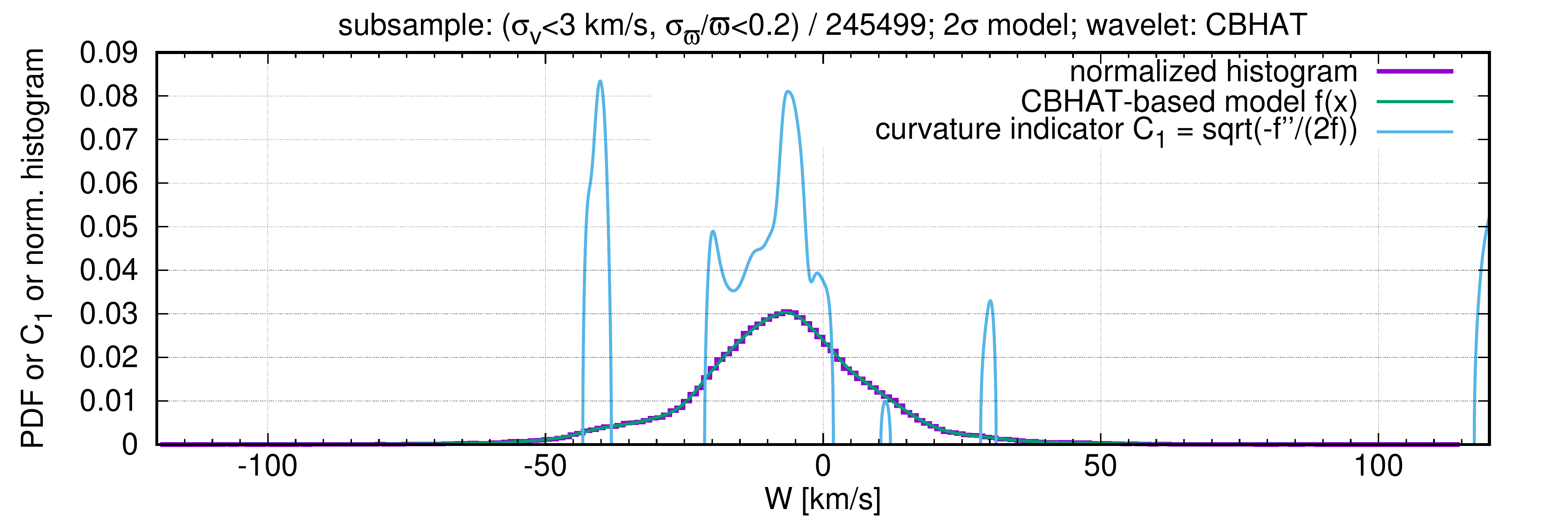}\\
\FigCap{\label{fig:1D-fater} Three univariate velocity PDFs constructed by the 1D
wavelet analysis for the larger sample ($\sigma_v<3$~km\,s$^{-1}$). Histograms normalized
by the bin width are shown and almost coincide with the PDF. The PDF curvature indicator
$C_1$ (see text) is also plotted.}
\end{figure}

We start from analysing 1D distributions for the $U$, $V$, and $W$ components of the
velocity. The corresponding 1D PDF models obtained for the $2$-sigma noise-cleaning level
are shown in Fig.~\ref{fig:1D-fater}, together with histograms. Additionally, we also plot
the so-called `contrasted curvature' of the PDF, defined as
\begin{equation}
C_1(x) = \sqrt{\frac{-f''(x)}{2f(x)}}.
\end{equation}
This entity is an $f''$-based indicator that (i) additionally highlights structures in the
distribution tails, where $f(x)$ is small, and (ii) has the same dimension as $f(x)$, so it
can be plotted on the same graph with $f(x)$.\footnote{This $C_1$ was derived from the
quadratic Taylor decomposition of $f(x)$, namely $1/C_1$ is equal to the abscissa shift
$\delta x$ necessary to reduce $f(x)$ to its local linear gradient.}

We can highlight the following conclusions from the 1D analysis.
\begin{enumerate}
\item The 1D PDFs are either unimodal or have rather inconclusive irregular shape (in case
of $V$).

\item Our $C_1$ indicator reveals several inhomogeneities related to the behaviour of the
PDF curvature. Formally, they represent some 1D density concentrations, but of course we do
not interpret them further because they likely involve multiple overlapped 2D or 3D
structures.

\item Three different noise thresholding levels (1,2,3-sigma) resulted in almost the same
PDF models (so we showed just the 2-sigma ones). This is expected, because whenever the
sample size is so large, the significance threshold becomes very abrupt: a particular
structure is either non-significant at all or quickly attains high significance. This
contrasts with what was obtained for exoplanetary samples \citep{Baluev18b}, because
exoplanetary samples were not so rich.

\item Apparently, all three PDF models follow the histograms very closely. One may ask
then: is the wavelet analysis indeed necessary here? Could we just stick with simple
histograms? Of course, we intentionally plotted such well-fitting histograms for
comparison. We could arbitrarily increase the bin width, loosing small-scale details, or
decrease it, resulting in an increased noise. Basically, the wavelet analysis adaptively
finds some most suitable local smoothing scale thanks to its multiresolution nature. Also,
the histogram says nothing reliable about the formal significance of PDF structures.

\item The standalone structure revealed by the $C_1$ indicator for $V=+75\ldots+110$~km/s
is an artifact, because it corresponds to the PDF level of $\sim 10^{-5}$, far in the tail
where there is no sample sources. Similar artifacts exist in other 1D distributions (partly
seen in the 1D W-plot, and beyond abscissa margins in the U-plot). These false structures
are likely analogous to the Gibbs phenomenon from the Fourier analysis \citep{Baluev18b}.
In this work they are very small in absolute magnitude and easily distingushable by
additional investigation. In 2D distributions they appear as a subtle diffraction-like ring
surrounding the entire sample.
\end{enumerate}

The 1D analysis is expectedly inconclusive due to the overlapping effect. We need to
consider at least 2D distributions to secure more reliable statistical groups.

\subsection{2D analysis}
Now let us proceed to the bivariate distributions. Considering three pairwise combinations
$(U,V)$, $(U,W)$, and $(V,W)$, we passed each sample through the 2D analysis pipeline from
\citep{BalRodShai19}. This resulted in 2D models of the corresponding PDFs. After that, we
also computed their Laplacians. This was done using the asymptotic
approximation~(\ref{laplacian}) for a small scale $a$.

After some experimenting with graphs we introduced, by analogy with the curvature indicator
$C_1$, an additional `log-balanced' curvature indicator $C_2'$:
\begin{equation}
C_2' = \log(1+\sigma^2 |C_2|)\, \mathrm{sign}\, C_2, \quad C_2=\frac{\Delta f}{2f}.
\end{equation}
Here $\sigma^2=\sigma_1^2+\sigma_2^2$ is the cumulative variance of two variables that we
consider, and $C_2$ is analogous to $C_1$. This new quantity $C_2'$ introduces a mixed
linear-logarithmic scale (linear for small $C_2$, but logarithmic for large $|C_2|$,
preserving its sign). This enables a more clear and contrast visual representation of
$\Delta f$ in a 2D heatmap plot.

\begin{figure}[tb]
\includegraphics[width=0.99\textwidth]{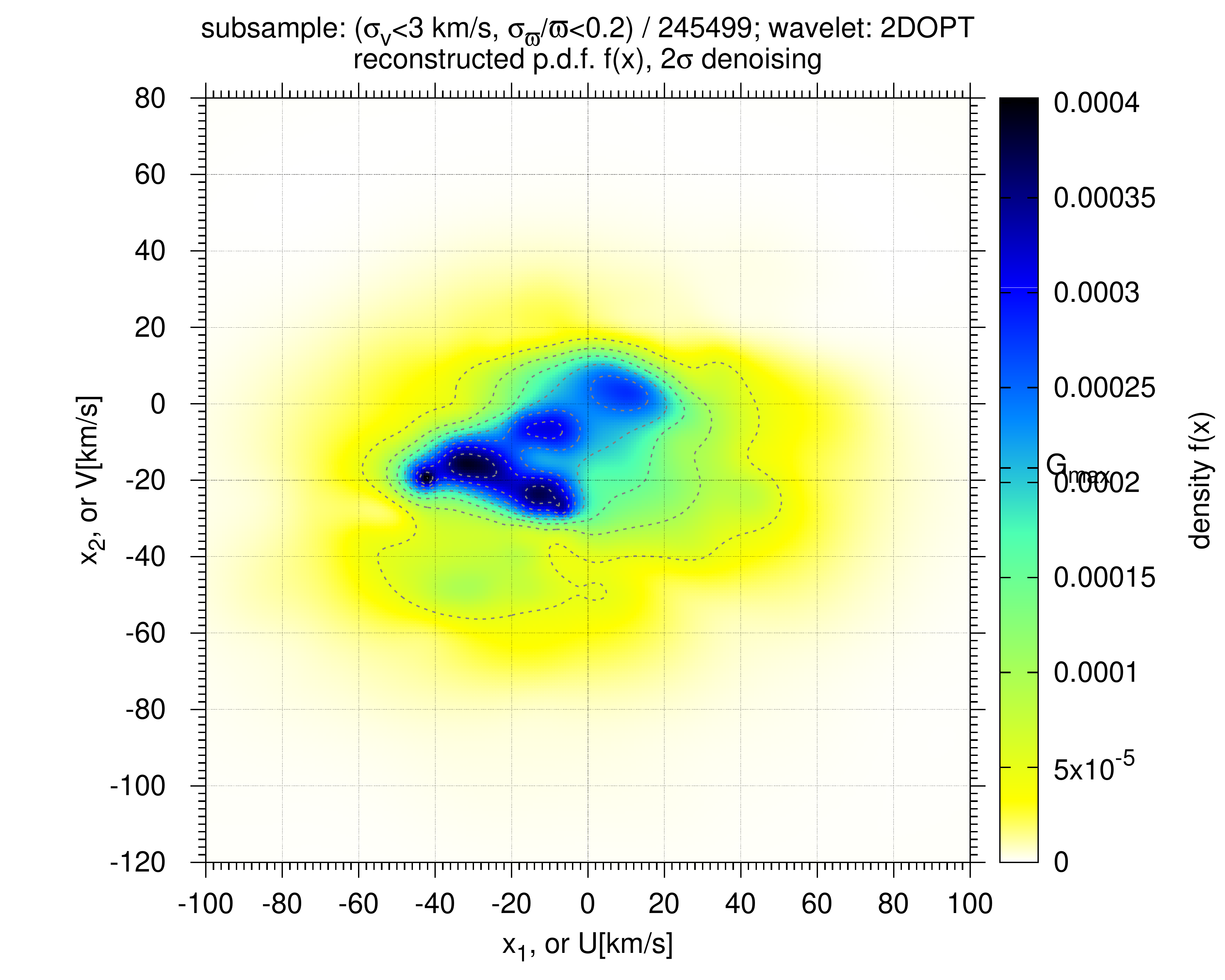}\\
\includegraphics[width=0.99\textwidth]{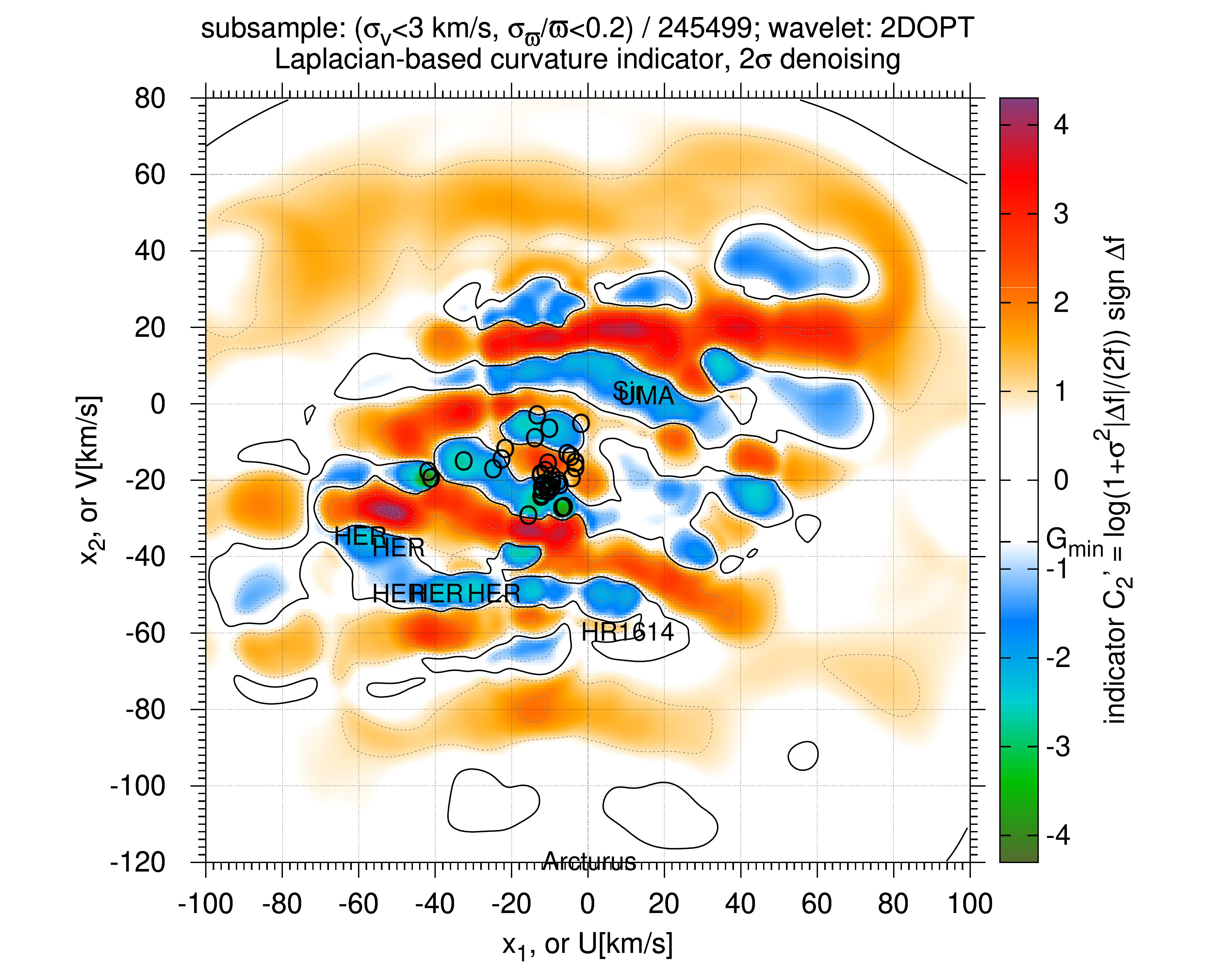}
\FigCap{\label{fig:2D-faterUV} Bivariate $(U,V)$-PDF constructed by the 2D wavelet analysis
for the larger sample ($\sigma_v<3$~km\,s$^{-1}$), and its Laplacian. Levels $G_{\max}$ in
PDF and $G_{\min}$ in the Laplacian correspond to the maximum of $f(x)$ for a
radially-symmetric 2D Gaussian with the same cumulative variance $\sigma^2$. Anything above
$G_{\max}$ (below $G_{\min}$) is more peaky than this Gaussian. The coloring of the
Laplacian map starts at $\pm |G_{\min}|$, and solid level curves follow $\Delta f\equiv 0$.
Known moving groups are also plotted as text labels or O-circles.}
\end{figure}

\begin{figure}[tb]
\includegraphics[width=0.99\textwidth]{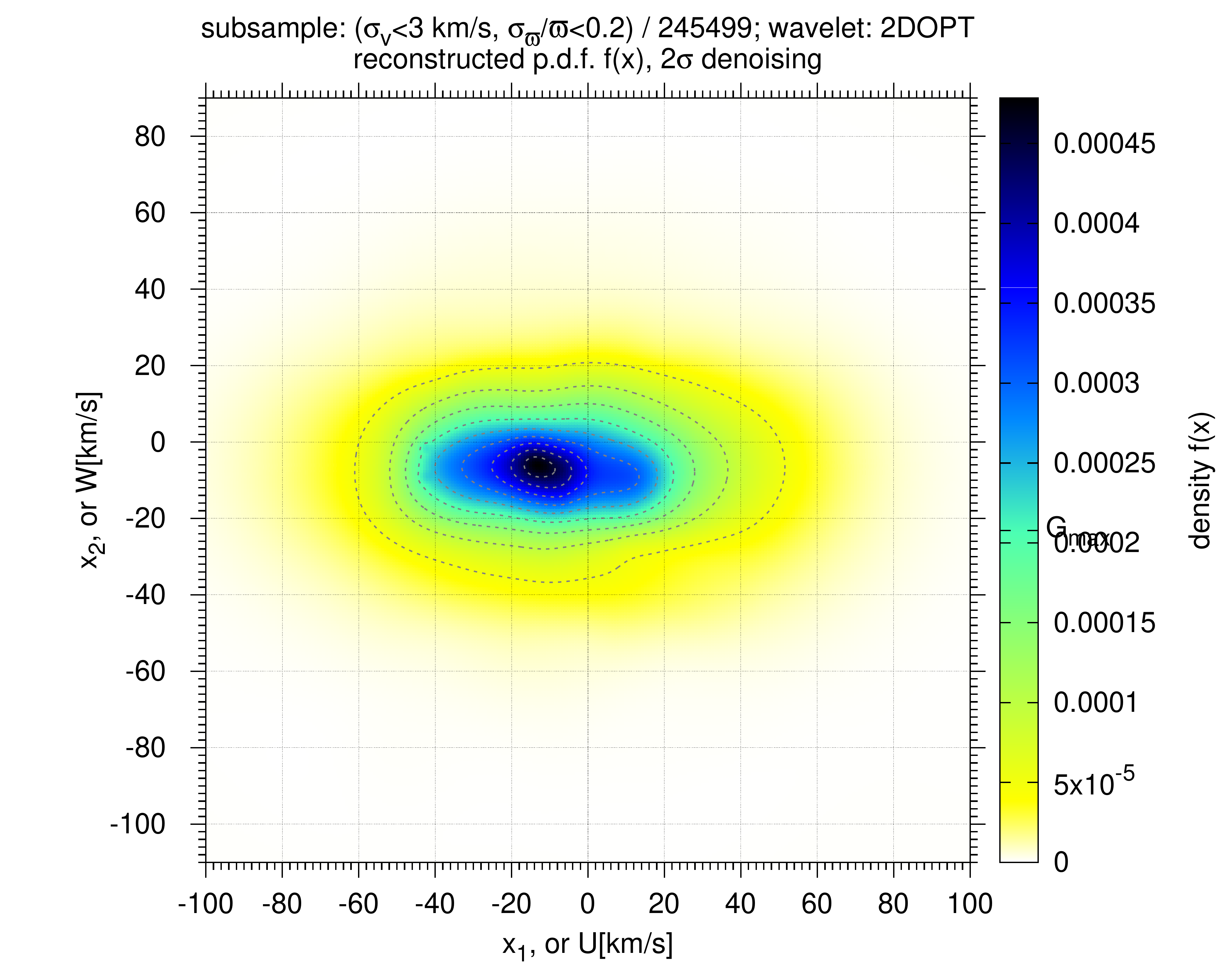}\\
\includegraphics[width=0.99\textwidth]{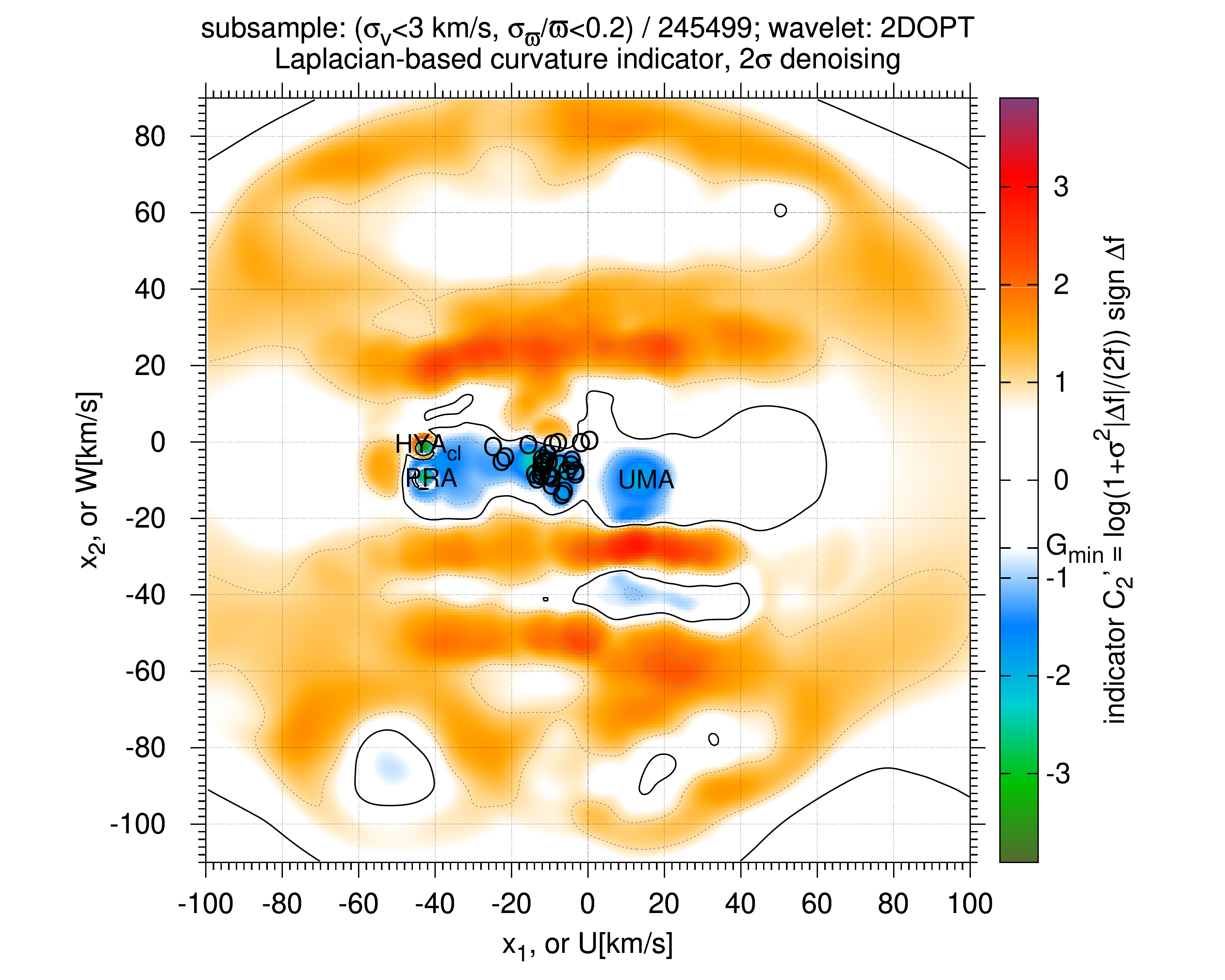}
\FigCap{\label{fig:2D-faterUW} Same as Fig.~\ref{fig:2D-faterUV} but for $(U,W)$
variables.}
\end{figure}

\begin{figure}[tb]
\includegraphics[width=0.99\textwidth]{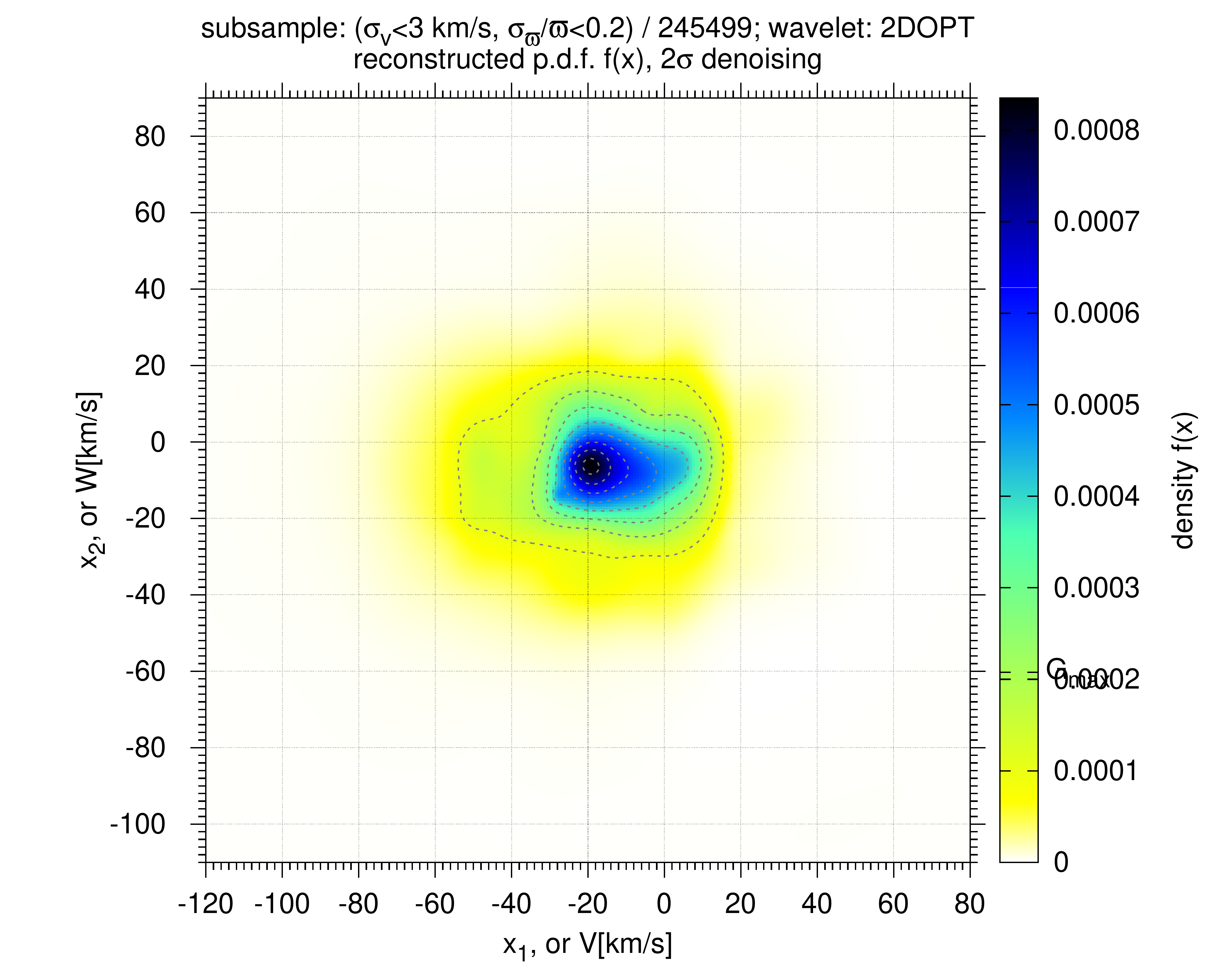}\\
\includegraphics[width=0.99\textwidth]{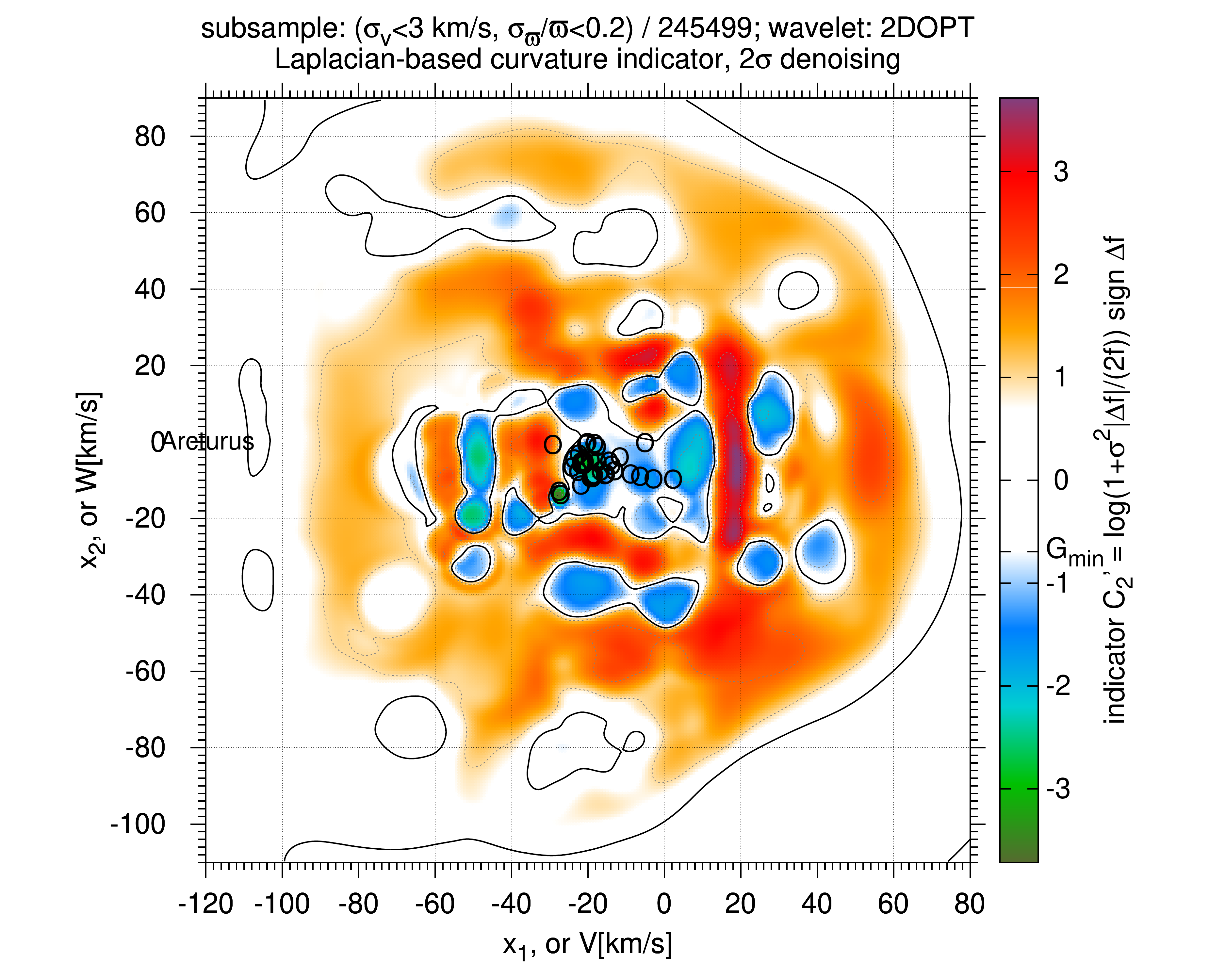}
\FigCap{\label{fig:2D-faterVW} Same as Fig.~\ref{fig:2D-faterUV} but for $(V,W)$
variables.}
\end{figure}

Our main results are shown in Fig.~\ref{fig:2D-faterUV}, Fig.~\ref{fig:2D-faterUW}, and
Fig.~\ref{fig:2D-faterVW}. We plot the PDF model itself (only for the two-sigma noise
threshold), and the corresponding Laplacian indicator $C_2$. In the latter case, we also
draw a zero-level contour line to outline possible isolated geometric structures, if any.
Finally, we also put in these plots all the known moving groups from
Appendix~\ref{sec_kmg}.

In the Laplacians we can see a lot of fine structures, standalone as well as irregular
patterns, likely representing partly overlapped configuration. The most rich plots appear
for $(U,V)$ and $(V,W)$ variables. Unfortunately, most of these 2D structures involve
overlapping effects, even if they appear as isolated at the first view. The overlapping
issue was not resolved entirely even in two dimensions, and it appeared here even more
complicated than it was for the asteroid families \citep{BalRod20}.

\begin{figure}[tb]
\includegraphics[width=0.99\textwidth]{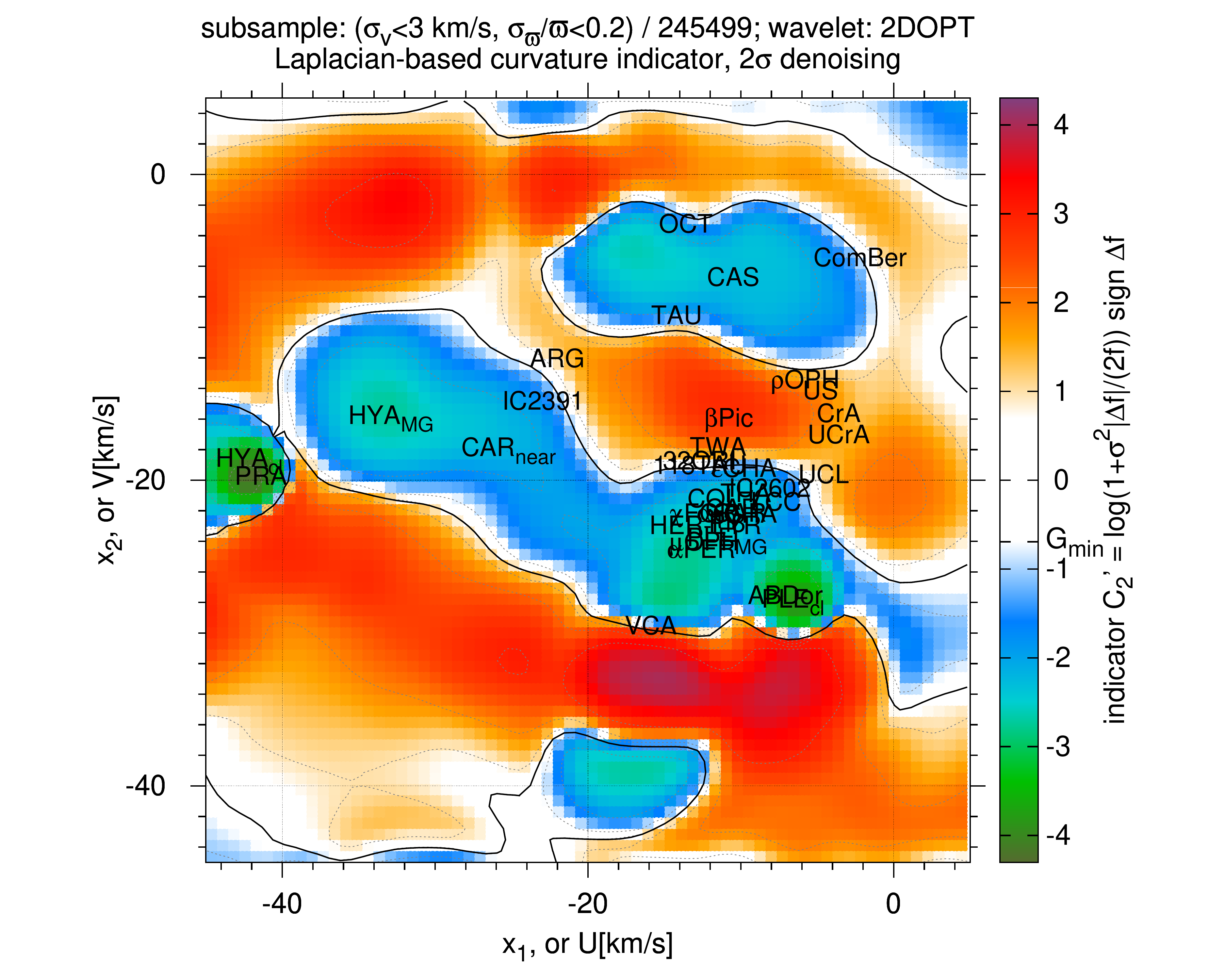}\\
\includegraphics[width=0.99\textwidth]{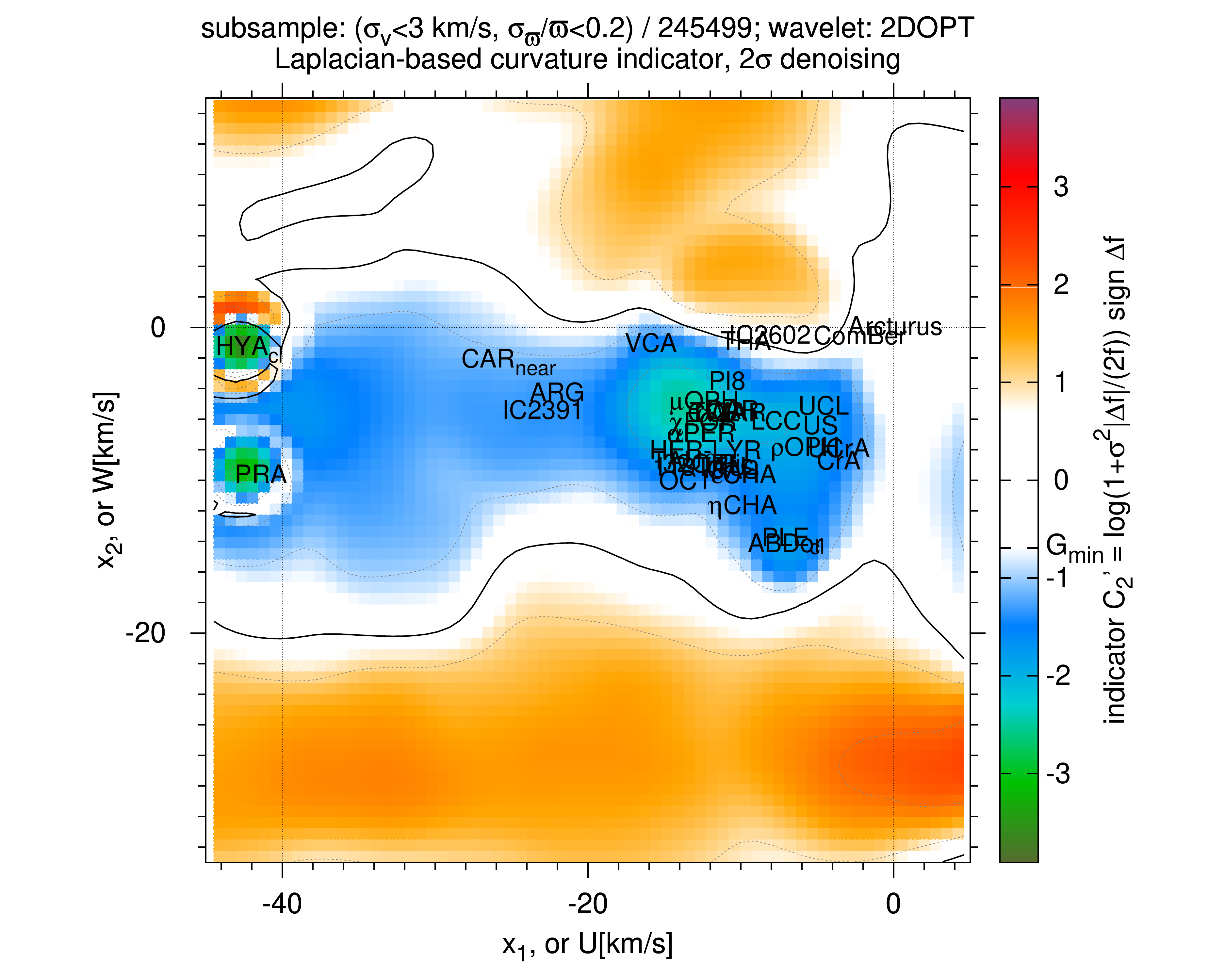}
\FigCap{\label{fig:2D-fater2UVUW} Central parts of Fig.~\ref{fig:2D-faterUV} and
Fig.~\ref{fig:2D-faterUW} expanded.}
\end{figure}

\begin{figure}[tb]
\includegraphics[width=0.99\textwidth]{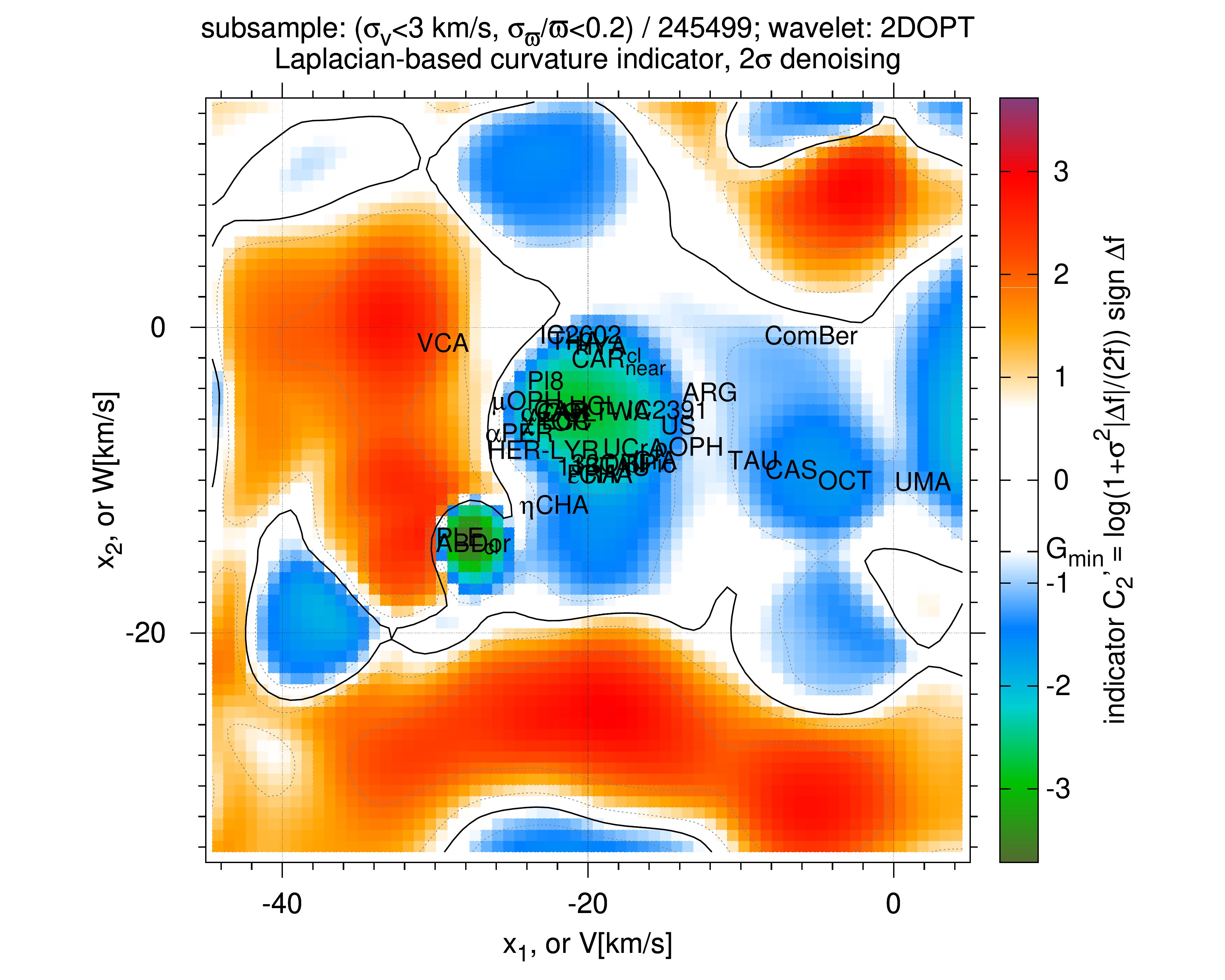}
\FigCap{\label{fig:2D-fater2VW} Central part of Fig.~\ref{fig:2D-faterVW} expanded.}
\end{figure}

Most of the known moving groups concentrate densely in the central part of the graphs. They
all mix into a few big diffuse maxima, so it does not seem there are good chances to
disentangle them in our work. The corresponding text labels cannot be resolved in these
plots, and to demonstrate the issue even further we give their expanded zooms in
Fig.~\ref{fig:2D-fater2UVUW} and Fig.~\ref{fig:2D-fater2VW}.

However, there are multiple peripherial PDF concentrations that may be cross-identified in
different 2D projections. They represent some new high-velocity moving groups that we may
now investigate and characterize.

\subsection{Disentangling standalone 3D moving groups}
We used the same approach to disentangle overlapped families as in \citep{BalRod20}. For
each isolated domain found in any of our 2D distributions, we cut out the corresponding
subsample. The subsample was cut out based on an elliptic mask selected manually to cover
the necessary 2D domain. Then for all these subsamples we performed 1D wavelet analysis
with respect to the remaining third variable. This enabled us to identify possible
overlapped components as 1D clusters located at different levels of the third coordinate.
Of course, this analysis did not always allow to perform an unambiguous disentangling. We
often faced an irresolvable issues related to too dense overlapping, or overlapping of more
than two groups within the same 2D domain, or difficulties due to complicated irregular 3D
geometry of the groups.

However, we were able to unambiguously disentangle almost $20$ moving groups listed in
Table~\ref{tab:wgroups}. In this table we give an approximate range in each dimension of
the UVW-space, and the 2D diagrams where the group appears. The boundaries of each cluster
correspond to the domain where $\Delta f<0$ (in the 2D projection that revealed this group)
or from the range where $f''<0$ (for dimensions analysed by the 1D algorithm). This means
that tails of each group are ignored. Virtually these ranges refer to only kernel portion
of a group. For example, if the group is radially symmetric and Gaussian then such a 2D
boundary would correspond to $\sqrt 2$-sigma deviation, while the 1D-boundary would be the
$1$-sigma one. However, most of these groups do not look to be either symmetric or
Gaussian, and under a close investigation their shape often appears rather complicated.
Moreover, boundaries of a small-scale group can be distorted (typically understated) due to
a large-scale background PDF curvature (see Fig.~\ref{fig:group}). In some uncertain cases
we had to determine the boundaries rather subjectively, but in general we expect that they
correspond to only the central part of each groups, covering very roughly about $\sim 50\%$
of group members.

From the other side, we could identify several famous open clusters (Praesepe, Hyades,
Pleiades and AB Doradus), for which our range estimates appear significantly larger than
those given in literature. We cannot confirm that narrow ranges based on our wavelet
analysis, though this might be an effect of velocity uncertainty ($1-3$~km/s for the most
sources), of the clusters internal structure or some other issue. Notice that e.g.
\citet{Riedel17} and \citet{Gagne18} provide very different velocity ranges for Hyades. We
worked only in the space UVW, so our velocity ranges may possibly include stars that
deviate a lot in term of XYZ. This may have an effect of additional velocity spreading.

Apart from those famous clusters, we could presumably identify in this list only the
Arcturus stream \citep{Navarro04}. Other our statistical groups do not have a
cross-identification with any of the known moving groups from the Appendix~\ref{sec_kmg},
so they represent some new ones.

\MakeTable{lcccccl}{\textwidth}{\label{tab:wgroups}Disentangled moving groups detected by bivariate wavelet analysis}
{\hline
No.& 2D diagrams & $U$, km/s & $V$, km/s & $W$, km/s & comment \\
\hline\\
1a & UV         & +35..+70   & +30..+50   & -25..+15  & might be joined into\\
1b & UV, VW     & -40..+30   & +20..+50   & -40..-20  & a single arc-like structure\\
\hline\\
2  & UV, VW     & -40..+0    & +20..+34   & -5..+20   & \\
\hline\\
3  & UV, VW     & -65..-40   & 0..+10     & -20..+15  & \\
\hline\\
4a & VW         & -45..+25   & -35..-10   & -45..-30  & might be joined\\
4b & VW, UV     & -30..+35   & -10..+10   & -38..-50  & into a single structure\\
\hline\\
5  & VW, UV     & -20..+20   & 0..+10     & +10..+25  & \\
\hline\\
6  & VW, UV     & -20..+20   & -10..0     & +10..+16  & \\
\hline\\
7  & UV, UW     & -45..-40   & -22..-17   & -12..-7   & Praesepe cl.\\
\hline\\
8  & UV, UW     & -45..-40   & -22..-17   & -3..+1    & Hyades cl. \\
\hline\\
9  & UV, VW     & -10..-3    & -30..-25   & -18..-12  & Pleiades cl., AB Doradus \\
\hline\\
10 & VW, UV     & -70..-50   & -32..-42   & -22..-10  & \\
\hline\\
11 & VW, UV     & -24..-13   & -44..-36   & -23..-12  & \\
\hline\\
12 & VW, UV     & +20..+26   & -44..-36   & -25..-10  & \\
\hline\\
13 & UV, VW     & -40..+20   & -75..-60   & -20..+10  & \\
\hline\\
14 & UV, VW     & -40..+40   & -120..-90  & -40..+20  & Arcturus stream? \\
\hline\\
15 & UV, VW     & -4..+24    & -40..-16   & +5..+15   & \\
\hline\\
16 & VW         & -50..+60   & -80..0     & +50..+70  & very marginal and uncertain\\
\hline\\
17 & UW, VW     & -70..+40   & -80..0     & -70..-90  & very marginal and uncertain\\
\hline}

We should also notice that our wavelet algorithm was tuned to detect patterns of
near-circular shape, due to the use of radially symmetric wavelets. However, unlike the
asteroid families \citep{BalRod20}, the stellar moving groups tend to have irregular shape
in the velocity space. Some structures in Table~\ref{tab:wgroups} appear disproportionally
elongated or compressed in a single dimension. They appear statistically significant only
when projected onto the compressed axes. However, they would not be detected at all if they
did not align so luckily with the coordinate axes. Most probably, such arbitrarily oriented
elongated structures also exist, but remained undetected in our analysis.

\subsection{Verifying the statistical reliability of the results}
\begin{figure}[tb]
\includegraphics[width=0.99\textwidth]{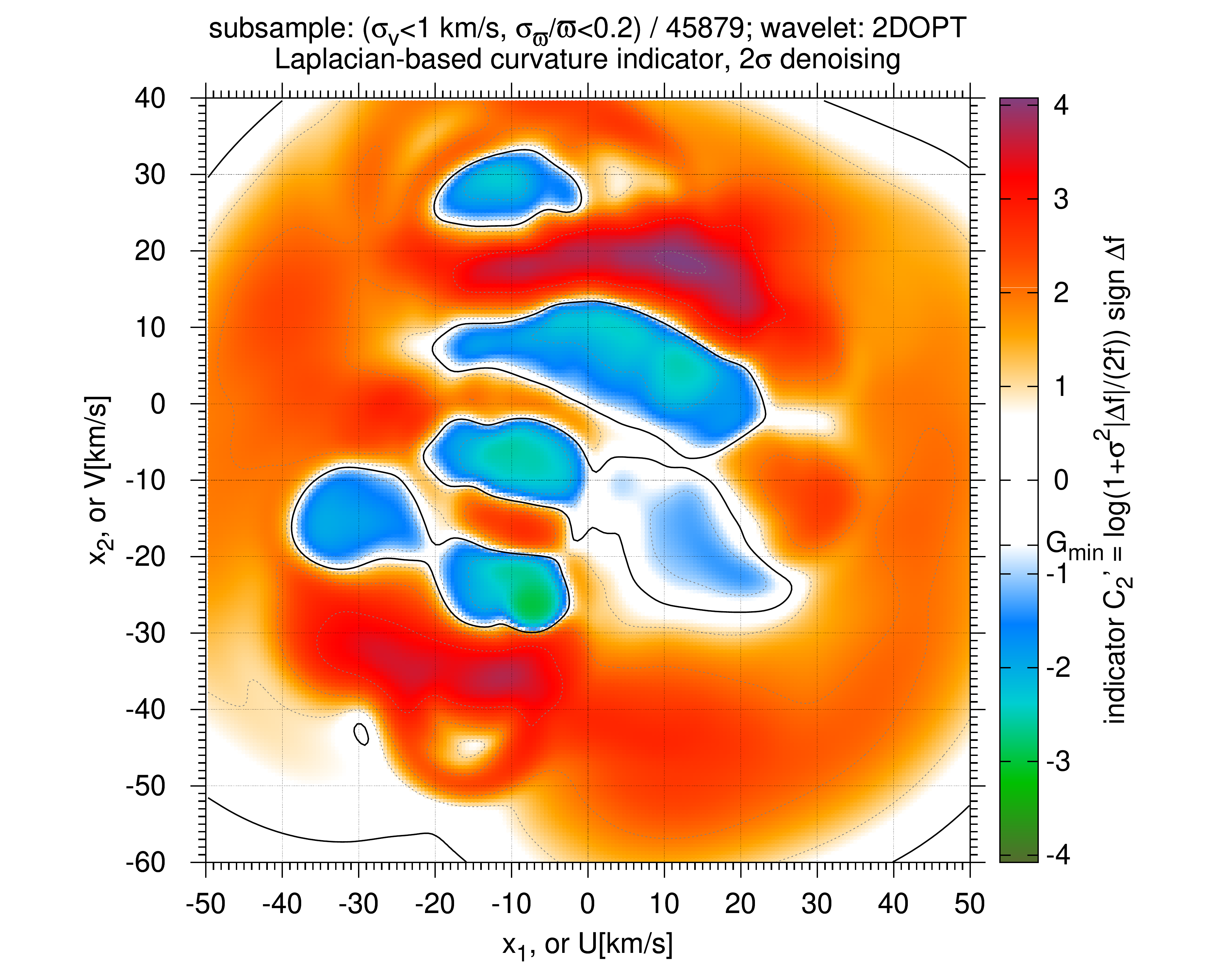}\\
\includegraphics[width=0.99\textwidth]{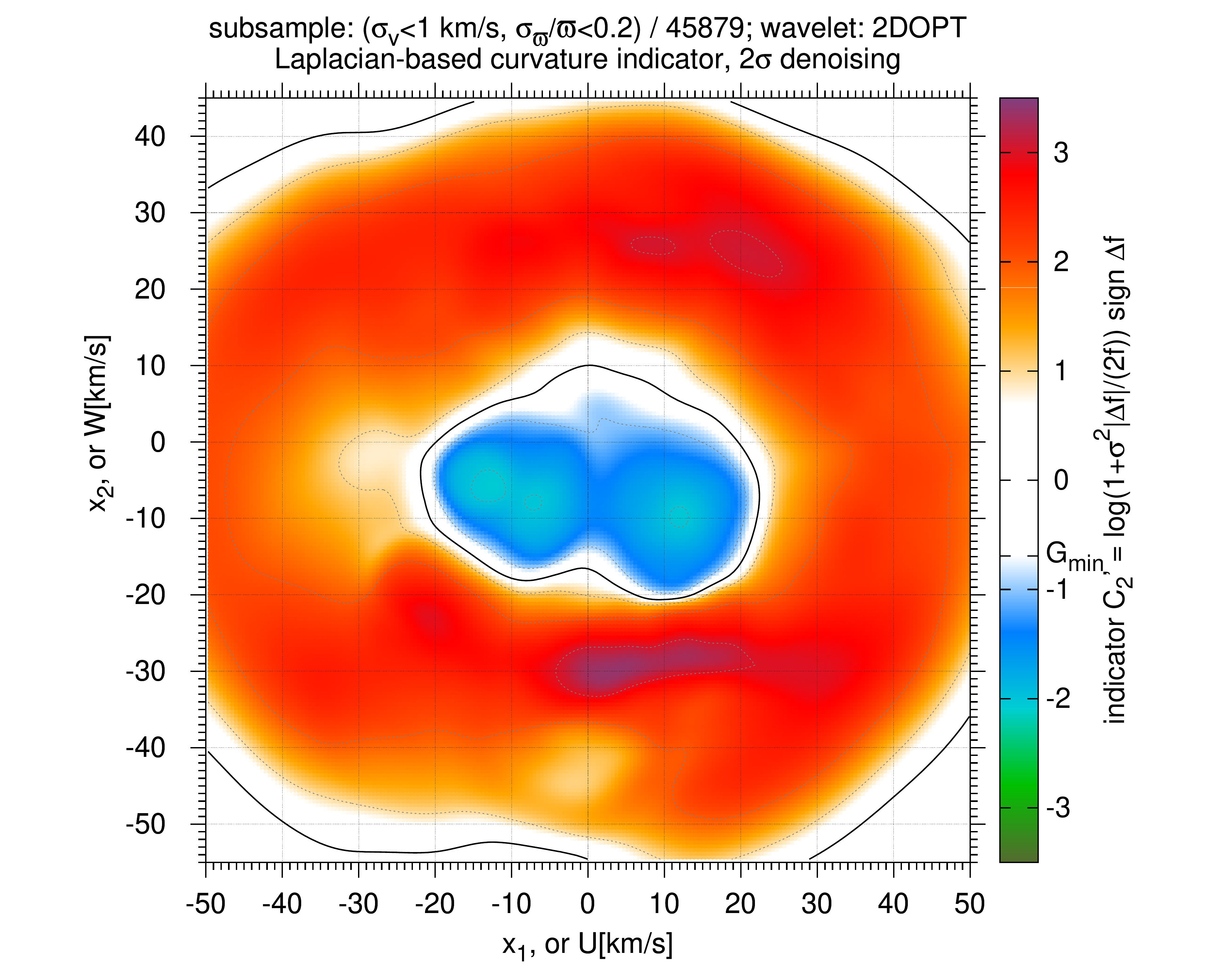}
\FigCap{\label{fig:2D-UVUW} Laplacians of the $(U,V)$ and $(U,W)$ PDFs reconstructed by 2D
wavelet analysis for the smaller sample ($\sigma_v<1$~km\,s$^{-1}$). See also notes from
Fig.~\ref{fig:2D-faterUV}. PDFs themselves are omitted, as well as moving groups labels.}
\end{figure}

\begin{figure}[tb]
\includegraphics[width=0.99\textwidth]{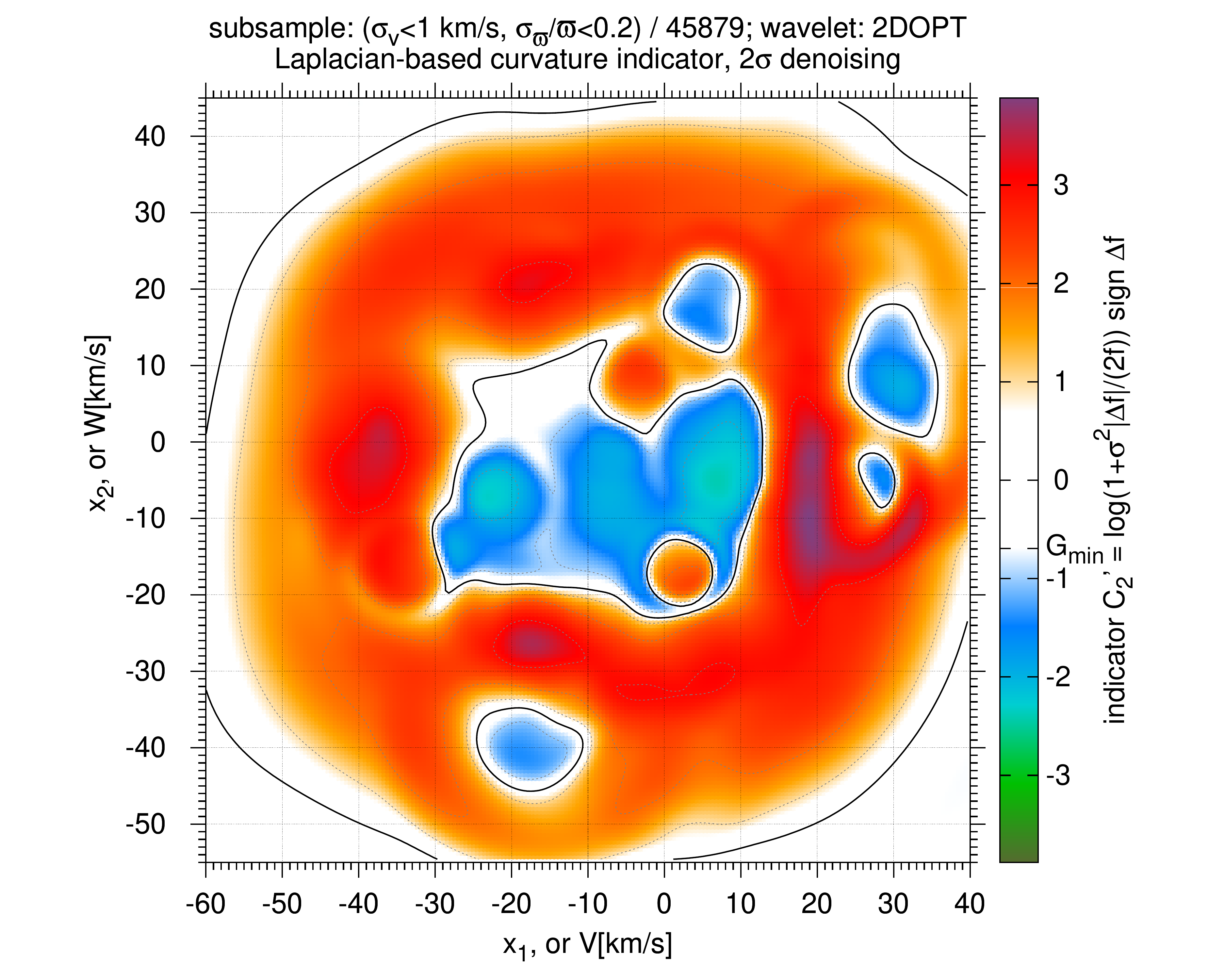}
\FigCap{\label{fig:2D-VW} Same as Fig.~\ref{fig:2D-UVUW} but for the $(V,W)$ pair.}
\end{figure}

Let us now try to verify how adequate was our wavelet algorithm in terms of noise
thresholding, i.e. does it indeed filter out all (or most) noisy patterns as expected. We
use the following approach for that: analyse two similar samples~--- a big one and a
smaller~--- and compare the results. We consider the analysis to be reliable in view of
this criterion, if all patterns revealed in the smaller sample are confirmed by the larger
one.

So, in addition to our primary sample, we also processed the smaller subsample with $N\sim
50000$ sources ($\sigma_v<1$~km\,s$^{-1}$). These results are presented in
Fig.~\ref{fig:2D-UVUW} and Fig.~\ref{fig:2D-VW}.

One might expect that the second sample should be capable to reveal finer structures,
because it contains more accurate velocities. But in actuality there is no such subtle
patterns, while the reduction of sample size increased statistical noise instead, so the
number of detectable structures decreased drastically. Nonetheless, we can see five
peripheral spots in these graphs indicating some statistically significant PDF convexities.
Each of them is confirmed by the larger-sample analysis of
Fig.~\ref{fig:2D-faterUV}--\ref{fig:2D-faterVW}. Neither of patterns detected in the small
sample is disproved by the larger one. Therefore, this suggests additional confirmation
that our analysis algorithm does not generate false structures.

\section{Conclusions}
Our underlying goal was to assess the applicability and usefullness of the wavelet analysis
in the stellar statistics. Although there were multiple interesting results like the list
of moving groups above with confirmed statistical significance, this work revealed several
important pitfalls.
\begin{enumerate}
\item There is a heavy overlapping effect between different stellar groups. In fact, nearly
all known moving groups merge so densely in the UVW-space that they form together just a
few large-scale maxima. Even though we tried to disentangle several groups in the 3D velocity
space UVW, it is still possible that some of them are heterogeneous when viewed in the
coordinate space XYZ. For example, the AB~Doradus moving group and the Pleiades cluster
have very similar velocity components, so they entirely overlap in the UVW-space. Although
there are clues that these two stellar associations may have common origin
\citep{Ortega07}, presently they are spatially separated from each other. In this work we
did not investigate the distribution in the XYZ space, so it may appear that some of the
moving groups listed in Table~\ref{tab:wgroups} should be split into independent
overlapping subgroups. The complete kinematical analysis should deal with the entire 6D
space UVWXYZ, which is currently far away from practical capabilities of our method. Its
extension from 2D to 6D would require significant additional theory work and qualitative
improvements to the computing speed.

\item The geometry of the stellar groups in the UVW-space looks quite complicated. They
often do not look like compact spherical or near-spherical density concentrations. Instead
of that, they often appear disproportionally elongated in one or two dimensions, and
possibly even curved. In some part this may be an effect of overlapping as well, but this
can also reflect some natural property coming from stellar dynamics. The wavelet analysis
technique has certain difficulties with processing such irregular patterns. Because of a
simplified radially symmetric wavelet shape, our analysis inherently tries to decompose any
elongated or membrane-like structure into a set of more or less oval-like `clouds' that may
need to be combined manually, e.g. as in \citep{Ramos18}. This issue can be partly
mitigated by using elliptically distorted wavelets through the use of additional scale
parameters, but this would again require a drastic increase of computing time due to larger
dimensions.

\item Justifying the statistical significance of the results is an important issue. We
again emphasize that to correctly threshold the sample noise effect, we should use the
extreme-value CWT distribution rather than the single-value one. The distribution of a
single wavelet coefficient (a particular value of the CWT function) is only useful if we
knew in advance which wavelet coefficient (at what shift-scale position) we plan to test.
However, in practice we usually test blindly $\sim 10^3$ or $\sim 10^4$ of almost
independent wavelet coefficients within a single CWT map. This eventually causes an effect
of p-hacking, because the number of false alarms then grows proportionally, and false
detections are then guaranteed even for very small p-value thresholds down to
$10^{-3}$-$10^{-4}$. This requires to either perform a correction for multiple testing, or
to use some aggregate cumulative statistic, e.g. the maximum CWT value as in our algorithm,
or a chi-square-like statistic as in \citep{McEwen04}.
\end{enumerate}

\Acknow{This work is supported by the Russian Foundation for Basic Research,
grant~17-02-00542~A. The authors thank the anonymous reviewer for providing useful
suggestions on this manuscript.}

\bibliographystyle{astron}
\bibliography{wavstars}

\appendix
\section{Known moving groups}
\label{sec_kmg}
In this appendix, we provide summary of stellar moving groups discussed in the literature
so far. They are given in Table~\ref{tab:mmg} and Table~\ref{tab:scmg}. They contain the
following data: velocity components $UVW$, the distance $D$, and the citation. All ranges
are those taken from the papers. We notice that the Herculis stream is actually a whole
standalone branch, possibly containing individual local subgroups \citep{Monari19}. So we
give only its rough UVW values in Table~\ref{tab:mmg}, and without distance. The
Scorpii--Centauri branch is also very big and likely contains multiple stellar groups
within, so we put them in a separate Table~\ref{tab:scmg}.
\newcommand{\mgname}[1]{\multicolumn{5}{l}{#1}}
\MakeTable{ccccl}{\textwidth}{\label{tab:mmg}Different known moving groups}
{
\hline
U [km/s] & V [km/s] & W [km/s] & D [pc] & Ref. \\
\hline\\[-0.5em]
\mgname{118 Tau}\\
$-12.8\pm2.1$ & $-19.1\pm2.8$ & $-9.2\pm1.6$ & ${\sim}102$ & \citet{Gagne18}\\[0.5em]
\mgname{32 Orinis (32 Ori)}\\
$-11.8\pm0.4$ & $-18.5\pm0.4$ & $-8.9\pm0.3$ & ${\sim}97$ & \citet{Riedel17}\\
$-12.8\pm2.2$ & $-18.8\pm2.2$  & $-9.0\pm2.0$ & ${\sim}95$ & \citet{Gagne18}\\
$-11.2\pm2.7$ & $-19.7\pm1.4$ & $-9.1\pm0.6$ & ${\sim}106$ & \citet{LeeSong19}\\[0.5em]
\mgname{$\beta$ Pictoris ($\beta$ PMG)}\\
$-10.1\pm2.1$ & $-15.9\pm0.8$ & $-9.2\pm1.0$ & $31\pm21$ & \citet{Torres08-book}\\
$-10.94\pm2.06$ & $-16.25\pm1.30$ & $-9.27\pm1.54$ & $9-73$ & \citet{Malo13}\\
$-11.03\pm1.38$ & $-15.61\pm1.72$ & $-9.24\pm2.50$ & $18-40$& \citet{Gagne14}\\
$-11.16\pm2.06$ & $-16.19\pm1.32$ & $-9.27\pm1.35$ & $9-73$& \citet{Malo14}\\
$-10.522\pm3.167$ & $-15.964\pm2.039$ & $-9.2\pm1.61$ & ${\sim}18$& \citet{Riedel17}\\
$-10.9\pm2.2$ & $-16.0\pm1.2$ & $-9.0\pm1.0$ & ${\sim}18$ & \citet{Gagne18}\\
$-10.0\pm2.6$ & $-16.2\pm1.4$ & $-8.9\pm1.4$  & ${\sim}24$ & \citet{LeeSong19}\\[0.5em]
\mgname{$\varepsilon$ Chamaeleontis ($\varepsilon$ CHA)}\\
$-11.0\pm1.2$ & $-19.9\pm1.2$ & $-10.4\pm1.6$ & $109\pm9$ & \citet{Torres08-book}\\ 
$-10.9\pm0.8$ & $-20.4\pm1.3$ & $-9.9\pm1.4$ & ${\sim}110$ & \citet{Riedel17}\\	
$-9.9\pm1.6$ & $-19.3\pm2.2$ & $-9.7\pm2.0$ & ${\sim}102$ & \citet{Gagne18}\\[0.5em]
\mgname{$\eta$ Chamaeleontis ($\eta$ CHA)}\\
$-10.2\pm0.2^*$  & $-20.7\pm0.1$ & $-11.2\pm0.1$ & ${\sim}94^*$& \citet{Riedel17}\\
$-10.0\pm1.6$ & $-22.3\pm2.8$ & $-11.7\pm1.8$ & ${\sim}95$& \citet{Gagne18}\\[0.5em]
\mgname{$\mu$ Ophiuchi / Mamajek2 ($\mu$ Oph)}\\
$-12.5$ & $-24.1$ & $-4.9$ & ${\sim}176$ & \citet{Jilinski09}\\[0.5em]
\mgname{$\rho$ Ophiuchi ($\rho$ Oph)}\\
$-5.9\pm1.3$ & $-13.5\pm4.7$ & $-7.9\pm4.3$ & ${\sim}131$ & \citet{Gagne18}\\[0.5em]
\mgname{$\chi^{01}$ Fornacis ($\chi^{01}$ For)}\\
$-12.29\pm0.98$ & $-20.95\pm0.92$ & $-4.9\pm1.07$ & ${\sim}99$ & \citet{Riedel17}\\
$-12.54\pm0.96$ & $-22.24\pm1.41$ & $-6.26\pm2.21$ & ${\sim}100$ & \citet{Gagne18}\\[0.5em]
\mgname{AB Doradus (ABDMG)}\\
$-6.8\pm1.3$ & $-27.2\pm1.2$& $-13.3\pm1.6$ & $34\pm26$ & \citet{Torres08-book}\\
$-7.12\pm1.39$ & $-27.31\pm1.31$ & $-13.81\pm2.16$ & $7-77$ & \citet{Malo13}\\
$-6.96\pm1.18$ & $-27.23\pm1.68$ & $-13.90\pm1.94$ & $19-50$ & \citet{Gagne14}\\
$-7.11\pm1.39$ & $-27.21\pm1.31$ & $-13.82\pm2.26$ & $11-64$ & \citet{Malo14}\\
$-7.031\pm2.136$ & $-27.241\pm1.929$ & $-13.983\pm1.859$ & ${\sim}18$ & \citet{Riedel17}\\
$-7.2\pm1.4$ & $-27.6\pm1.0$ & $-14.2\pm1.8$ & ${\sim}14$& \citet{Gagne18}\\
$-7.3\pm4.6$ & $-27.4\pm2.6$ & $-13.6\pm1.9$& ${\sim}22$ & \citet{LeeSong19}\\[0.5em]
\mgname{Arcturus}\\
${\sim}0\pm50$ & ${\sim}{-120}\pm20$ & ${\sim0}\pm25$ & & \citet{Navarro04}\\[0.5em]
\hline
}
\addtocounter{table}{-1}\MakeTable{ccccl}{\textwidth}{Different known moving groups. Continue}
{
\hline
U, km/s & V, km/s & W, km/s & D, pc & Ref. \\
\hline\\[-0.5em]
\mgname{Argus / Carinae--Velorum (ARG)}\\
$-22.0\pm0.3$ & $-14.4\pm1.3$ & $-5.0\pm1.3$ &$106\pm51$ & \citet{Torres08-book}\\
$-21.78\pm1.32$ & $-12.08\pm1.97$ & $-4.52\pm0.50$ & $8-68$ & \citet{Malo13}\\
$-21.54\pm0.87$ & $-12.24\pm1.67$ & $-4.63\pm2.74$ & $15-48$ & \citet{Gagne14}\\
$-21.78\pm1.32$ & $-12.08\pm1.97$ & $-4.52\pm0.50$ & $8-68$ & \citet{Malo14}\\
$-22.133\pm1.992$ & $-12.122\pm1.755$ & $-4.324\pm0.774$ & ${\sim}27$ & \citet{Riedel17}\\
$-23.4\pm4.9$ & $-14.0\pm2.6$ & $-4.9\pm1.7$ & ${\sim}80$ & \citet{LeeSong19}\\[0.5em]
\mgname{Carinae (CAR)}\\
$-10.2\pm0.4$ & $-23.0\pm0.8$ & $-4.4\pm1.5$ & $85\pm35$& \citet{Torres08-book}\\
$-10.50\pm0.99$ & $-22.36\pm0.55$ & $-5.84\pm0.14$ & $46-88$ & \citet{Malo13}\\
$-10.50\pm0.99$ & $-22.36\pm0.55$ & $-5.84\pm0.14$ & $46-88$ & \citet{Malo14}\\
$-10.691\pm1.763$ & $-22.582\pm0.532$ & $-5.746\pm0.178$ & ${\sim}72$ & \citet{Riedel17}\\
$-10.66\pm0.67$ & $-21.92\pm1.02$& $-5.48\pm1.01$ & ${\sim}53$&\citet{Gagne18}\\
$-10.6\pm1.5$ & $-22.5\pm1.2$ & $-4.0\pm0.9$ & ${\sim}117$ & \citet{LeeSong19}\\[0.5em]
\mgname{Carinae-Near}\\
$-27.020\pm3.044$ & $-18.255\pm1.819$ & $-3.021\pm1.147$ & ${\sim}20$ & \citet{Riedel17}\\
$-25.3\pm3.2$ & $-18.1\pm1.9$  & $-2.3\pm2.0$ & ${\sim}28$ &\citet{Gagne18}\\[0.5em]
\mgname{Castor (CAS)}\\
$-10.7$ & $-8$ & $-9.7$ & $<45$ & \citet{Montes01} \\
$-10.6\pm3.7$ & $-6.8\pm2.3$ & $-9.4\pm2.1$ & ${\sim}20-50$ & \citet{Ribas03}\\[0.5em]
\mgname{Columbae (COL)}\\
$-13.2\pm1.3$ & $-21.8\pm0.8$ & $-5.9\pm1.2$ & $82\pm30$ & \citet{Torres08-book}\\
$-12.24\pm1.03$ & $-21.32\pm1.18$ & $-5.58\pm0.89$ & $35-81$ & \citet{Malo13}\\
$-12.14\pm0.51$ & $-21.29\pm1.27$ & $-5.61\pm1.69$ & $26-63$ & \citet{Gagne14}\\
$-12.24\pm1.08$  & $-21.27\pm1.22$ & $-5.56\pm0.94$ & $35-81$ & \citet{Malo14}\\
$-12.311\pm2.321$ & $-21.681\pm1.43$ & $-5.694\pm1.322$ & ${\sim}49$ & \citet{Riedel17}\\
$-11.90\pm1.04$ & $-21.28\pm1.29$ & $-5.66\pm0.75$ & ${\sim}42$ & \citet{Gagne18}\\
$-12.9\pm2.1$ & $-21.5\pm1.2$ & $-5.0\pm1.1$ & ${\sim}75$ & \citet{LeeSong19}\\[0.5em]
\mgname{Comae Berenices (Coma Ber)}\\
$-2.512\pm1.868$ & $-5.417\pm1.364$ & $-1.204\pm1.876$ & ${\sim}88$ & \citet{Riedel17}\\
$-2.30\pm0.53$ & $-5.51\pm0.44$ & $-0.61\pm0.71$ & ${\sim}85$ & \citet{Gagne18}\\[0.5em]
\mgname{Coronae Australis (CRA)}\\
$-3.7\pm1.3$ & $-15.7\pm2.2$ & $-8.8\pm2.2$ & ${\sim}140$ & \citet{Gagne18}\\[0.5em]
\mgname{Herculis (Her)}\\
 ${\sim}-25$ & ${\sim}-50$ & & & \citet{Monari19}, end of ridge\\
 ${\sim}-40$ & ${\sim}-50$ & & & \citet{HuntBovy18}\\
 ${\sim}-50$ & ${\sim}-38$ & & & \citet{Gaia18kin}, 1st branch\\
 ${\sim}-50$ & ${\sim}-50$ & & & \citet{Gaia18kin}, 2nd branch\\
 ${\sim}-60$ & ${\sim}-35$ & & & \citet{Skuljan99}\\[0.5em]
\mgname{Herculis--Lyrae (Her--Lyr)}\\
$-15.4$  & $-23.4$ & ${\sim}{-12}$ & $25$ & \citet{LSant06}\\
$-12.41\pm3.725 $ & $-23.03\pm3.59$ & $-8.11\pm3.80$ & & \citet{Eisenbeiss13}\\[0.5em]
\hline
}
\addtocounter{table}{-1}\MakeTable{ccccl}{\textwidth}{Different known moving groups. Continue}
{
\hline
U, km/s & V, km/s & W, km/s & D, pc & Ref. \\
\hline\\[-0.5em]
\mgname{Hyades cluster (Hya)}\\
$-41.1\pm0.23$ & $-19.2\pm0.23$ & $-1.4\pm0.23$ & ${\sim}46$ & \citet{Riedel17}\\
$-42.27\pm2.01$ & $-18.79\pm0.94$ & $-1.47\pm1.10$ & ${\sim}42$ & \citet{Gagne18}\\
${\sim}-42$ & ${\sim}-19$ & & & \citet{Gaia18kin}\\[0.5em]
\mgname{Hyades moving group}\\
$-33$  & $-16$ &  & & \citet{Gaia18kin}\\[0.5em]
\mgname{IC 2391 super cluster (IC 2391)}\\
$-20.6$ & $-15.7$ & $-9.1$ & ${\sim}155$ & \citet{ByN99}\\
$-23.04\pm1.10$ & $-14.89\pm3.40$ & $-5.48\pm0.78$ & ${\sim}149$ & \citet{Gagne18}\\
$-23.63$ & $-14.45$ & $-5.53$ & & \citet{Postnikova19}\\[0.5em]
\mgname{IC 2391 stream}\\
$-21.11$ & $-7.21$ & $-6.65$ & & \citet{Postnikova19}\\[0.5em]
\mgname{IC 2602}\\
$-8.22\pm 1.18$ & $-20.60\pm2.61$ & $-0.58\pm 0.65$ & ${\sim}146$ & \citet{Gagne18}\\[0.5em]
\mgname{Octantis (OCT)}\\
$-14.5\pm0.9$ & $-3.6\pm1.6$ & $-11.2\pm1.4$ & $141\pm34$ & \citet{Torres08-book}\\
$-13.673\pm1.749$ & $-4.8\pm1.678$ & $-10.927\pm1.029$ & ${\sim}115$ & \citet{Riedel17}\\
$-13.7\pm2.4$ & $-3.3\pm1.3$ & $-10.1\pm1.4$ & ${\sim}114$ & \citet{Gagne18}\\[0.5em]
\mgname{Pl8}\\
$-11.01\pm1.15$ & $-22.89\pm1.96$ & $-3.59\pm0.74$ & ${\sim}126$ & \citet{Gagne18}\\[0.5em]
\mgname{Pleiades cluster (Ple)}\\
$-6.7\pm1.7$ & $-28.0\pm1.8$ & $-14.0\pm1.2$ & ${\sim}134$& \citet{Gagne18}\\
${\sim}-7$ & ${\sim}-28$ & & & \citet{Gaia18kin}\\[0.5em]
\mgname{Pleiades moving group}\\
$-11$  & $-24$ &  & & \citet{Gaia18kin}\\[0.5em] 
\mgname{Praesepe cluster (PRA)}\\
$-41.5\pm0.9$ & $-19.8\pm0.5$ & $-9.7\pm1.1$ & ${\sim}182$  & \citet{vanLeeuwen09}\\[0.5em]
\mgname{Sirius moving group}\\
$10$ & $3$ & & & \citet{Gaia18kin}\\[0.5em] 
\mgname{TAU}\\
$-14.3\pm3.1$ & $-9.3\pm4.5$ & $-8.8\pm3.4$ & ${\sim}122$ & \citet{Gagne18}\\[0.5em]
\mgname{Tucanae--Horologii (THA)}\\
$-9.9\pm1.5$& $-20.9\pm0.8$ & $-1.4\pm0.9$ & $48\pm7$ & \citet{Torres08-book}\\
$-9.88\pm1.51$ & $-20.70\pm1.87$ & $-0.90\pm1.31$ & $28-92$ & \citet{Malo13}\\
$-9.70\pm1.05$ & $-20.47\pm1.68$ & $-0.78\pm2.38$ & $38-51$& \citet{Gagne14}\\
$-9.93\pm1.55$ & $-20.72\pm1.79$ & $-0.89\pm1.41$ & $36-71$& \citet{Malo14}\\
$-10.6\pm0.2$ & $-21.0\pm0.2$ & $-2.1\pm0.2$ & ${\sim}40$& \citet{Kraus14} \\
$-9.802\pm4.01$ & $-20.883\pm2.883$ & $-1.023\pm1.458$ & ${\sim}40$& \citet{Riedel17}\\
$-9.79\pm0.87$ & $-20.94\pm0.79$ & $-0.99\pm0.72$ & ${\sim}42$  & \citet{Gagne18}\\
$-9.6\pm1.7$ & $-21.0\pm1.1$ & $-1.0\pm0.6$ & ${\sim}43$ & \citet{LeeSong19}\\[0.5em]
\hline
}
\addtocounter{table}{-1}\MakeTable{ccccl}{\textwidth}{Different known moving groups. Continue}
{
\hline
U [km/s] & V [km/s] & W [km/s] & D [pc] & Ref. \\
\hline\\[-0.5em]
\mgname{TW Hydrae (TWA)}\\
$-10.5\pm0.9$ & $-18.0\pm1.5$ & $-4.9\pm0.9$ & $48\pm13$& \citet{Torres08-book}\\ 
$-9.87\pm4.15$ & $-18.06\pm1.44$ & $-4.52\pm2.80$ & $28-92$& \citet{Malo13}\\
$-11.12\pm0.90$ & $-18.88\pm1.56$ & $-5.63\pm2.78$ & $40-62$ & \citet{Gagne14}\\
$-10.53\pm3.50$ & $-18.27\pm1.17$ & $-5.00\pm2.15$ & $49-92$& \citet{Malo14}\\
$-11.7\pm0.9$ & $-17.3\pm1.3$ & $-5.0\pm1.0$ & ${\sim}61$ & \citet{Ducourant14}\\
$-10.954\pm3.043$ & $-18.036\pm2.332$ & $-4.846\pm1.703$ & ${\sim}58$& \citet{Riedel17}\\
$-11.6\pm1.8$ & $-17.9\pm1.8$ & $-5.6\pm1.6$ & ${\sim}55$& \citet{Gagne18}\\
$-12.2\pm2.6$ & $-18.7\pm1.2$ & $-6.0\pm0.9$ & ${\sim}56$ & \citet{LeeSong19}\\[0.5em]
\mgname{Upper Coronae Australis (UCRA)}\\
$-3.7\pm3.0$  & $-17.1\pm1.8$ & $-8.0\pm1.2$ & ${\sim}147$ & \citet{Gagne18}\\[0.5em]
\mgname{Ursa Majoris (UMA)}\\
$14.278\pm2.64$ & $2.392\pm0.594$  & $-8.987\pm0.407$ & ${\sim}25$& \citet{Riedel17}\\
$14.8\pm1.0$ & $1.8\pm1.2$ & $-10.2\pm2.6$ & ${\sim}25$& \citet{Gagne18}\\[0.5em]
\mgname{Volantis--Carinae (VCA)}\\
$-16.0\pm5.4$ & $-29.6\pm0.9$ & $-1.1\pm0.6$ & ${\sim}87$& \citet{LeeSong19}\\[0.5em]
\hline
}

\MakeTable{ccccl}{\textwidth}{\label{tab:scmg}Scorpii--Centauri moving groups}
{
\hline
U, km/s & V, km/s & W, km/s & D, pc & Ref. \\
\hline\\[-0.5em]
\mgname{$\alpha$ Persei cluster ($\alpha$ Per)}\\
$-11$ & $-26$ & $-7$  & $176$  & \citet{deZeeuw99}\\
$-12.7\pm0.9$&$-24.6\pm0.7$&$-7.0\pm0.2$& ${\sim}173$ & \citet{vanLeeuwen09}\\[0.5em]
\mgname{Cas-Tau}\\
$-13$ &$-20$ & $-6$ & $140$ & \citet{deZeeuw99}\\[0.5em]
\mgname{Cep OB2}\\
$7$ & $2$ & $-1$ & $657$ & \citet{deZeeuw99}\\[0.5em]
\mgname{Cep OB6}\\
$-14$ & $-29$ & $-6$ & $243$ & \citet{deZeeuw99}\\[0.5em]
\mgname{Col 121}\\
$-13$ & $-4$& $-4$ & $546$ & \citet{deZeeuw99}\\[0.5em]
\mgname{Lac OB1}\\
$5$ & $-7$ & $-4$ & $373$ & \citet{deZeeuw99}\\[0.5em]
\mgname{Lower Centauri--Crucis (LCC)}\\
$-12$ & $-13$ & $-7$ & $118$ & \citet{deZeeuw99}\\
$-7.8\pm2.7$ & $-21.5\pm3.8$ & $-6.2\pm1.8$ & ${\sim}109$ & \citet{Gagne18}\\[0.5em]
\mgname{Per OB2}\\
$1$ & $-11$ & $-1$ & $310$& \citet{deZeeuw99}\\[0.5em]
\mgname{Tr 10}\\
$-24$ & $-24$ & $-8$ & $362$ & \citet{deZeeuw99}\\[0.5em]
\mgname{Upper Centauri--Lupi (UCL)}\\
$-9$ & $-17$ & $-6$ & $140$ & \citet{deZeeuw99}\\
$-4.7\pm3.8$ & $-19.7\pm3.0$ & $-5.2\pm1.7$ & ${\sim}126$ & \citet{Gagne18}\\[0.5em]
\mgname{Upper Scorpii (US)}\\
$-1$ & $-16$ & $-5$ & $145$ & \citet{deZeeuw99}\\
$-4.9\pm3.7$ & $-14.2\pm3.2$ & $-6.5\pm2.3$ & ${\sim}132$& \citet{Gagne18}\\[0.5em]
\mgname{Vel OB2}\\
$-16$ & $6$ & $-1$ & $415$ & \citet{deZeeuw99}\\[0.5em]
\hline
}

\end{document}